\documentclass[12pt]{article}
\usepackage{amsmath}
\usepackage{graphicx}
\usepackage{enumerate}
\usepackage{natbib}
\usepackage{url} 
\usepackage{bbm}
\usepackage{bm}
\usepackage{amssymb}
\usepackage{caption}
\usepackage{subcaption}
\usepackage{float}
\usepackage{placeins}

\usepackage{tikz} 
\usetikzlibrary{shapes,arrows}
\usetikzlibrary{automata,positioning}
\tikzstyle{block} = [draw, fill=white, rectangle, 
    minimum height=3em, minimum width=7em]
\tikzstyle{input} = [coordinate]
\tikzstyle{output} = [coordinate]

\usepackage{xcolor}
\definecolor{darkgreen}{rgb}{0.0, 0.5, 0.0}
\usepackage[linesnumbered,ruled,vlined]{algorithm2e}

\SetCommentSty{mycommfont}

\usepackage{xcite}
\usepackage{xr}
\makeatletter
\newcommand*{\addFileDependency}[1]{
  \typeout{(#1)}
  \@addtofilelist{#1}
  \IfFileExists{#1}{}{\typeout{No file #1.}}
}
\makeatother

\newcommand*{\myexternaldocument}[1]{%
    \externaldocument{#1}%
    \addFileDependency{#1.tex}%
    \addFileDependency{#1.aux}%
}

\myexternaldocument{appendices}

\newcommand{\blind}{1}

\begin{document}

\def\spacingset#1{\renewcommand{\baselinestretch}%
{#1}\small\normalsize} \spacingset{1}


\if1\blind
{
  \title{\bf Nonparametric monitoring of sunspot number observations: a case study}
  \author{Sophie Mathieu \\
    ISBA/LIDAM, UCLouvain\\
    and \\
		Laure Lef{\`e}vre \\
    Solar Physics and Space Weather department, Royal Observatory of Belgium \\
		 and \\
    Rainer von Sachs \\
    ISBA/LIDAM, UCLouvain \\
		    and \\
    V{\'e}ronique Delouille \\
    Solar Physics and Space Weather department, Royal Observatory of Belgium\\
		    and \\
    Christian Ritter \\
    ISBA/LIDAM, UCLouvain \\
		 and \\
		Fr{\'e}d{\'e}ric Clette \\
    Solar Physics and Space Weather department, Royal Observatory of Belgium}
  \maketitle
} \fi

\if0\blind
{
  \bigskip
  \bigskip
  \bigskip
  \begin{center}
    {\LARGE\bf Nonparametric monitoring of sunspot number observations: a case study}
\end{center}
  \medskip
} \fi

\bigskip
\begin{abstract}
Solar activity is an important driver of long-term climate trends and must be accounted for in climate models. Unfortunately, direct measurements of this quantity over long periods do not exist. The only observation related to solar activity whose records reach back to the seventeenth century are sunspots.
Surprisingly, determining the number of sunspots consistently over time has remained until today a challenging statistical problem. It arises from the need of consolidating data from multiple observing stations around the world in a context of low signal-to-noise ratios, non-stationarity, missing data, non-standard distributions and many kinds of errors. The data from some stations experience therefore severe and various deviations over time.\\
In this paper, we propose the first systematic and thorough statistical approach for monitoring these complex and important series. It consists of three steps essential for successful treatment of the data: smoothing on multiple timescales, monitoring using block bootstrap calibrated CUSUM charts and classifying of out-of-control situations by support vector techniques. \\ 
This approach allows us to detect a wide range of anomalies (such as sudden jumps or more progressive drifts), unseen in previous analyses.
It helps us to identify the causes of major deviations, which are often observer or equipment related. Their detection and identification will contribute to improve future observations. Their elimination or correction in past data will lead to a more precise reconstruction of the world reference index for solar activity: the International Sunspot Number. 

\end{abstract}

\noindent%
\vfill

\newpage
\spacingset{1} 

\section{Introduction}
\label{sec:intro}

The main objective of this paper is to provide an efficient monitoring for the sunspot numbers. Until recently, such a monitoring could not be developed with the available statistical methods due to the complexity of the series. 
Moreover, existing modeling and error quantification of the data were lacking. Advances in statistical process control for panel data combined with a thorough uncertainty model for the sunspot numbers~\citep{Mathieu2019} allow us to construct an appropriate monitoring method for these data.  \\
Our developed method bears however the potential to be applied to a much wider range of problems.
Hence, this paper follows two threads, one related to the specific patterns and considerations of sunspot observations, the other to the development, selection, and parametrization of a statistical methodology for monitoring them. 


\subsection{Overview}
\label{sec:overview}
Sunspots appear as dark areas on the Sun. They correspond to regions with a high local magnetic field and relate to the solar activity. 
The International Sunspot Number is an indicator of this activity. It is composed of the number of sunspots, $N_s$ and the number of sunspot groups, $N_g$ via a composite, $N_c = N_s + 10 N_g$.
This index is one of the most intensely used time-series in astrophysics \citep{Hathaway2010}. It enters into models of the influence of the solar irradiance on the Earth climate~\citep{Haigh2002, Ermolli2013} and in space weather predictions \citep{Temmer2001, Wang2014}. \\ 

Although astronomers started observing sunspots in the beginning of the seventeenth century, it remains surprisingly difficult to arrive at an accurate daily determination of their numbers. Three difficulties stand out: observability, resolution and interpretation. For Earth-based observatories, the Sun cannot be observed when there are clouds. Instruments with different resolutions may give rise to different counts of sunspots. Distinguishing sunspots and groups of sunspots require experience and even experts sometimes disagree. 
The estimate is therefore a weighted consensus among all observatories, also called ``stations'' participating in the effort. 
Different observers may vary in skill and their skills may vary over time.  Some stations observe systematically more sunspots than others and there may be shifts due to instrument changes.  Moreover, the observed series from different stations cover different time periods.
We are therefore facing panel data with missing values of various patterns, time-varying biases and non-overlapping time periods. \\
There are also intrinsic sources of variability. Some sunspots are only visible over short periods. Their number may therefore change during the day. 
Finally, the sunspot activity itself is subject to substantial variability at multiple time-scales. \\
The statistical challenge is to untangle observation-related variability from actual variations of the solar activity: to detect and correct the former and to extract the latter. 
These variability patterns are non-normal and autocorrelation naturally occurs. \\

There are also many disturbances such as jumps, drifts or oscillations.
Those have been partially studied on the short-term in the previous works by \cite{Morfill1991}, \cite{Vigouroux1994} and later \cite{Dudok2016}. More recently, \cite{Mathieu2019} developed a comprehensive uncertainty model that involves three types of station-dependent errors, at short-term, long-term time scales as well as during solar minima.   
It reveals the various irregularities that stations may exhibit. \\ 
Assuring the quality of the consolidated series therefore calls for an automated tool for supervising the observations in quasi real-time. This procedure should monitor the stations and send alerts when they start deviating to prevent the occurrence of large drifts. Owing to methodological advances in the sunspot numbers uncertainty modeling and in statistical process control (SPC), it is now possible to develop such a method.

\subsection{Related work}
\label{sec:existing}

Different procedures have been developed in the SPC literature to monitor complex panel of data.
The aim of this case-study is not to compare them. It is to adapt the most-promising existing procedure to efficiently control the deviations of the sunspot numbers over time.  
Therefore, we propose in the following a method based the dynamic screening system developed by \cite{Qiu2014}. \\ 
The method of \cite{Qiu2014} is based on extensions of the classical CUSUM~\citep{Page1961} chart. It is composed of two steps. First, the regular patterns (i.e. the mean and the variance) of the data are estimated on a subset of non-deviating or in-control (IC) series. Second, the data are standardized by these patterns and monitored by a CUSUM chart designed by a block bootstrap method. This procedure constructs, without any parametric assumption, a control scheme that is valid for non-normally distributed and serially correlated data. 
Moreover, the regular patterns of the data are estimated over a subset of stable series. 
This allows the chart to detect shifts in the mean level of each series, where the series is allowed to have a mean changing over time. Hence, the method can accommodate the quasi-periodic variations in the sunspot number series which are related to the eleven year solar cycle~\citep{Hathaway2010}. This cycle is intrinsic to the signal and unlinked to the long-term stability of the observations. 
The control scheme of \cite{Qiu2014} seems therefore particularly appropriate for our problem. It allows us to treat the sunspot data which are non-stationary, non-normally distributed and autocorrelated.  \\ 

In the following, we use and extend the work of \cite{Qiu2014}, to bridge the gaps between the method and the specific requirements of our problem. Those gaps are two-fold.
(1) The method of \cite{Qiu2014} ---as all other methods that we encountered in the literature--- cannot be used however without knowing a priori which stations are in-control. This information is not available for our data, where \emph{all} stations are expected to contain several kinds of deviations (jumps, oscillating shifts, etc) in their observation period. 
(2) The method operates with a control chart which sends an alert when a deviation is detected, yet without providing any information about the nature of the shift. Such information is however crucial for us, since it allows to further investigate the causes of the shifts. 
Although several methods have been developed to automatically predict the size of a shift after an alert (see for instance~\cite{Cheng2011} and the references therein), they are not adapted to data which are (simultaneously) non-normally distributed, serially correlated and contaminated by strong noise.  

\subsection{Aims}
\label{sec:aims}

In the following, we propose a nonparametric monitoring that is tailored to the complex features of the sunspot numbers: (a) the missing values, (b) the strong noise, (c) the complex autocorrelation structure and (d) the non-normality. Our method extensively exploits the information contained in the panel 
 to establish a robust IC reference from the network. 
This allows us to monitor the stations without prior information on their stability. 
We complete the method by a support vector machine (SVM) procedure that efficiently predicts the size and the shape of a shift once an alert has been raised. 
Although we could manually build a library with typical shapes and sizes to be compared to the deviations, we select the automatic SVM approach instead. 
 \\ 
The control scheme is then applied on past observations to study the deviations of the sunspot numbers. 
The procedure automatically detects major deviations identified recently by hand in some stations. It also unravels many other deviations, unseen in previous analyzes. In particular, small and persistent shifts that are difficult to identify manually are detected by the method. The precise information about the deviations predicted by the SVM procedures allows us to determine the causes of some prominent deviations. This sets the ground for a future enhancement of the quality of the series. 
Moreover, the monitoring procedure provides the possibility to be used in real-time to preserve the long-term stability of the stations. It also paves the way to a future redefinition of the International Sunspot Number based on several stations that are stable over time. \\

This article is structured as follows. In Section \ref{sec:data}, we present the main properties of the data and their model.
The methods are explained in Section \ref{sec:methods}. This includes the complete monitoring scheme as well as the SVM procedures to predict the size and shape of the deviations.  
In Section \ref{sec:results}, we apply the proposed method on the sunspot data at different scales, in order to detect both high- and low- frequency shifts and discuss the results on actual stations.
In a final section \ref{sec:conclusion}, we give some concluding remarks and perspectives. 
Supplementary materials provide some more details about the monitoring scheme as well as more examples of monitored stations.

\section{Data}
\label{sec:data}

The data and their specific features are first presented in this section. Then, we introduce the uncertainty model associated to the sunspot numbers. The component of the model that will be monitored in this paper is finally presented alongside with its estimating procedure. 

\subsection{Presentation of the dataset}

The period under study embraces the most recent part of the series and extends from January 1, 1981 (when the sunspot numbers production center moved from Zurich to Brussels) till December 31, 2019. 
It ranges from the descending phase of solar cycle (SC) 21 to the minimum between SC 24 and 25\footnote{\url{https://en.wikipedia.org/wiki/List_of_solar_cycles}} and covers thus three complete solar cycles. 
The data are composed of the daily observations of a network of 278 Earth-ground observatories disseminated across the world. The records contain the number of spots $N_s$, groups $N_g$ and composite $N_c$. They are distributed through the World Data Center Sunspots Index and Long-term Solar Observations (WDC-SILSO)\footnote{ The data are available at the following link: \url{http://www.sidc.be/silso/}}.
In the following, we denote by $t$,  $t \in 1,...,T$, the date-time of the observation and represent the index of the stations by $i$, $i \in 1,..., N=278$. 

The data have complex features that are described below. Those should be taken into account in the design of the monitoring procedure. 
\begin{enumerate}
\setlength{\itemsep}{0.1cm}
\item As studied in \cite{Mathieu2019} and previous works, the data are by nature non-normally distributed. 
\item 
The data contain around 70\% of missing values over the period studied. Those are mainly caused by the non-overlapping observing periods of the stations. Indeed, some stations started observing only recently while older stations stopped their activity well before 2019. The stations also contain various percentages of missing values (ranging from 15\% to 75\%) over their active observing period, mainly due to weather conditions that prevent the observation of the Sun.
\item The stations are also correlated across the panel and along time since a sunspot may stay from several minutes up to several months on the solar surface. 
\item Due to the observing conditions and the solar variability, the series experience a wide range of deviations that vary in shape and size. 
\end{enumerate}

\subsection{Model}
Let $Y_i(t)$ represent either the number of spots, groups or composite observed in station $i$ at time $t$.
The observations may be decomposed into a common solar signal, generically denoted by $s(t)$, corrupted by three types of station-dependent errors in a \emph{multiplicative} framework~\citep{Mathieu2019}: 
\begin{equation} 
\begin{split}
Y_i(t) = \left\{ \begin{array}{ll} (\epsilon_1(i,t)+\epsilon_2(i,t)+h(i,t))s(t) & \text{if} \  s(t)>0 \\ 
\epsilon_3(i,t) & \text{if} \  s(t)=0. \\ \end{array} \right.
\end{split}
 \label{E:model}
 \end{equation}
$s(t)$ is a latent variable representing the actual number of spots, groups or composite of the Sun. It cannot be directly observed but its mean may be estimated based on the observations of the network and later used as a proxy for $s(t)$.
$\epsilon_1$ denotes the short-term error representing counting errors and variable seeing conditions. We assume that $\mathbb{E}(\epsilon_1(i,t))=0$ where $\mathbb{E}$ denotes the expectation sign. Its variance and tail, that also vary in time, are studied extensively in \cite{Mathieu2019}. 
$\epsilon_2$ represents a long-term error. We are interested in estimating and monitoring its mean, denoted by $\mu_2(i,t)$, which represents the bias of the stations. At a later stage, we may also study the variance of $\epsilon_2$. 
We also introduce the random variable $h(i,t)$, where $h$ is a function of time which varies at a slower pace than $\epsilon_2$. This quantity may be associated to the background level of the stations (accounting e.g. for differences of instruments or counting methodologies of the stations) and is not a target of our monitoring procedures.   
The errors $\epsilon_1$, $\epsilon_2$ and $h$ vary on different time-scales. 
They affect $s(t)$ in a multiplicative way~\citep{Oh2012}.
Finally, $\epsilon_3$ captures effects like short-duration sunspots during solar minima (a detailed study about the distribution of $\epsilon_3$ can be found in \cite{Mathieu2019}). \\
The random variables $\epsilon_1$, $\epsilon_2$, $\epsilon_3$ and $h$ are assumed to be continuous and $\epsilon_1$, $\epsilon_2$, $\epsilon_3$, $h$ and $s(t)$ to be jointly independent.  Note that $\epsilon_1$, $\epsilon_2$ and $\epsilon_3$ would be equal to zero and $h$ be equal to one for a station that would be --- in absence of any measurement errors ---perfectly aligned with the solar signal. 

\subsection{Long-term bias}
\label{sec:bias}

As mentioned earlier, we are not so much interested in the short term error $\epsilon_1$ and our monitoring aims mostly at the component $\epsilon_2$ and more specifically its mean  $\mu_2$. The even longer term levels of the stations expressed by $h$ are simply estimated and removed from the data.
To this end, we isolate the long-term error from the other components of the model in the step-wise approach described below.
We first divide the observations by scaling factors to roughly compensate for different observing conditions: $Z_i(t) = \frac{Y_i(t)}{\kappa_i(t)}$. These piece-wise constant scaling factors $\kappa_i(t)$ are computed as the slope of the ordinary least-squares regression between the observations of the stations and the median of the observations ($\underset{1 \leq i\leq N}{\text{med}} Y_i(t)$) on periods of 8 months for $N_s$, 14 months for $N_g$ and 10 months for $N_c$. These values are selected by a statistical-driven study based on the Kruskal-Wallis test \citep{Kruskal_Wallis} that is completely described in the section 6.3 of~\cite{Mathieu2019}. \\
Afterward, we compute $M_t$, a robust proxy for $s(t)$ based on the median of the rescaled observations:
\begin{equation}
M_t = \underset{1 \leq i\leq N}{\text{med}} Z_i(t).
\label{E:Mt}
\end{equation}
Motivated from (\ref{E:model}), the observations $Y$ are then divided by $M_t$ to remove the main influence of the solar signal. They are also smoothed by a moving-average (MA) filter, represented by a $\star$ in the following equation. This smoothing process untangles $eh$, $eh = \epsilon_2 + h$, from the short-term error $\epsilon_1$:
\begin{equation}
\widehat{eh}(i,t) = \left( \frac{ Y_i(t)}{M_t} \right)^\star ~ \ \text{when} \ M_t>0,
\label{E:e2}
\end{equation}
where $\widehat{eh}$ denotes the estimator of the mean of $eh$, which is used as a proxy for $eh$. 
\begin{figure}[htb]
	\centering
		\includegraphics[scale=0.5]{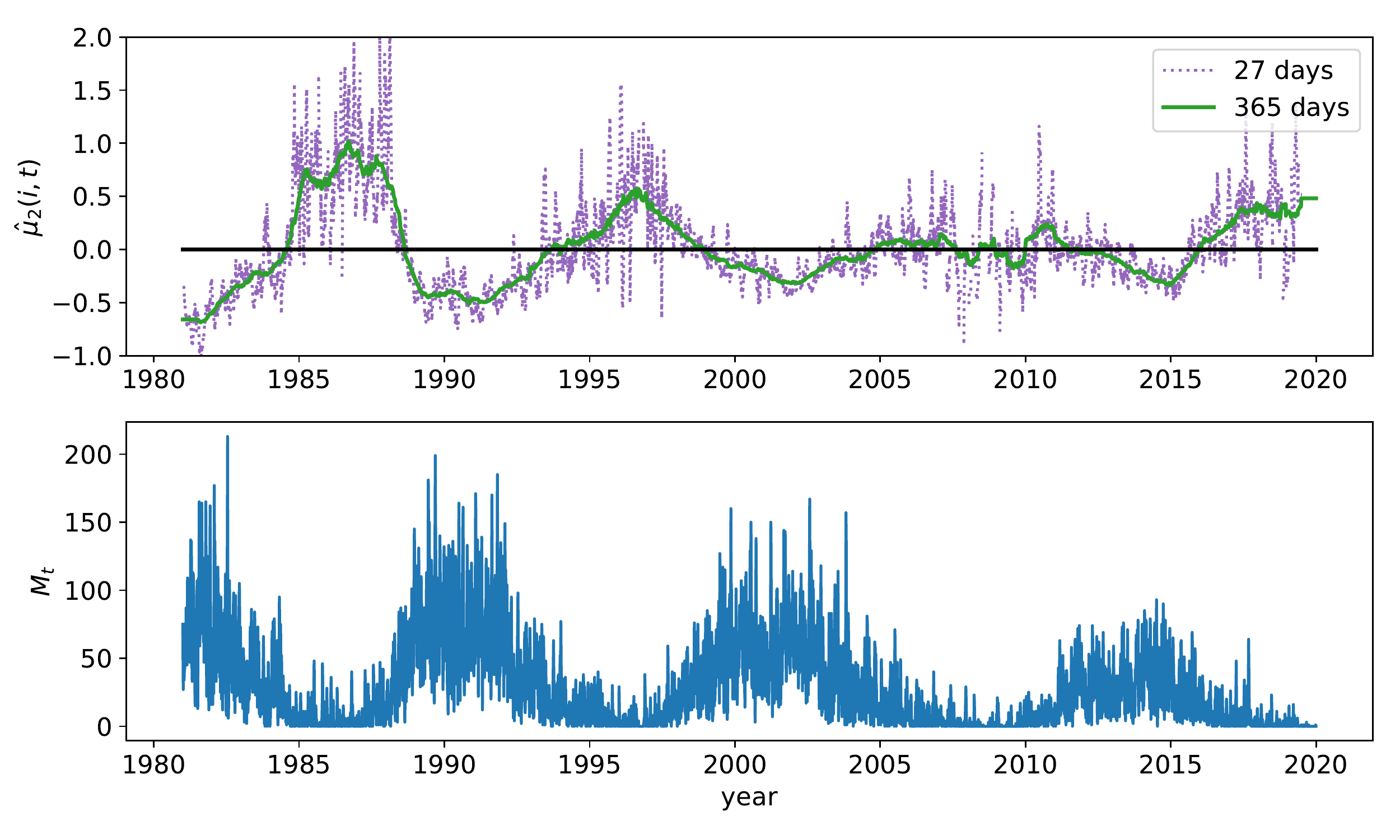}
		\caption{\footnotesize{Long-term bias, $\hat \mu_2(i,t)$ for $N_s$, in the station Locarno (Switzerland) over its observing period. The $\hat \mu_2$s are smoothed on 27 days (dotted line) to allow the detection of high-frequency shifts and smoothed on 365 days (plain line) to emphasize the low-frequency deviations. $M_t$ is also represented in the lower plot as an estimation of the actual number of spots. This figure clearly shows the eleven-year solar cycle~\citep{Hathaway2010} that is intrinsic to the signal. }}
	\label{fig:cycle}
\end{figure}
To analyze the various deviations of the data, different MA-filter window lengths may be used in (\ref{E:e2}). 
The low-frequency shifts such as persisting drifts are first studied at a yearly scale (i.e. with a window length of 365 days). Then, a window of length equal to 27 days will also be used to examine the high-frequency deviations such as sudden jumps. This value of 27 days corresponds to a physical scale of the data: one solar rotation. It appears to be sufficiently high to overcome the effects of the short-term regime, as demonstrated in~\cite{Mathieu2019}.
Finally, the levels of the stations are separated from the long-term bias by applying once again a MA smoothing process denoted by $\star\star$: 
\begin{equation}
\label{E:level}
\begin{split}
\hat \mu_2(i,t)=\widehat{eh}(i,t) - \widehat{eh}^{\star \star}(i,t),
\end{split}
\end{equation}
where the MA-filter window length should be larger than those of (\ref{E:e2}). 
It is selected here at eleven years or one solar cycle, a physical value that is larger than the time-scales of the long-term error $\epsilon_2$ considered here. It also seems appropriate since the location of the observatories or their telescope are unlikely to change much over time. \\
As an illustration, the $\hat \mu_2$s smoothed on 27 and 365 days are represented in Figure~\ref{fig:cycle} for a particular station. 
Since the $\hat \mu_2$s are a (smoothed) ratio, they depend on the level of the solar signal. This is clearly shown by comparing the lower and upper plots in the figure. 

\section{Ingredients of the method}
\label{sec:methods}

In this section, the complete monitoring procedure depicted in Figure~\ref{Fig:method} is explained. It is intentionally presented in a generic framework to allow the application of the method on the number of spots $N_s$, groups $N_g$ as well as composites $N_c$. 
There are three phases: (I) estimation of the in-control (IC) parameters of the data, (II) construction and use of the monitoring procedure and (III) identification of out-of-control patterns. \\
Phase I contains two steps. At first a subset of stations is selected from the panel which follows closely the median signal $M_t$ mentioned in Section \ref{sec:data}. This pool of stations is then used as a proxy for IC series in the nomenclature of \cite{Qiu2014}. They are used to determine the IC patterns (mean and variance) of the data and to provide the basis of the block-bootstrap procedure in Phase II.  After standardizing all series by the IC patterns, the CUSUM control chart is calibrated in phase II by a block bootstrap procedure from the pool of IC series. The scheme is then applied to the data for the monitoring. 
In Phase III, support vector machine (SVM) procedures predict the shifts size and shape on sub-series detected as out-of-control by the CUSUM, for easier problem diagnostic.


\begin{centering}
\begin{figure}[hbt]
\begin{tikzpicture}[auto, node distance=4.5cm, >=latex']

   \node (input) [block, text width=2.5cm, align=center, dashed, label=Ia] {Selection of IC processes};
    \node (fft) [block, right of=input, text width=2.5cm, align=center, label=Ib] {Estimation of IC patterns};
    \node (ths) [block, right of=fft, text width=3cm, align=center, label=II] {Design of the chart and monitoring of standardized data}; 
    \node (iasc) [block, right of=ths, text width=2.6cm, align=center, dashed, label=III] {Estimation of shifts size/shape by SVMs};
      
     \draw [->] (input) -- node[midway, above] {} (fft);
     \draw [->] (fft) -- node[name=u3] {} (ths);
     \draw [->] (ths) -- node[name=u5] {} (iasc);
		

\end{tikzpicture}
\caption{\footnotesize{Pipeline of the procedure. It is in particular in the dashed blocks that we contribute with completely new ingredients of the method. 
}}
 \label{Fig:method}
\end{figure}
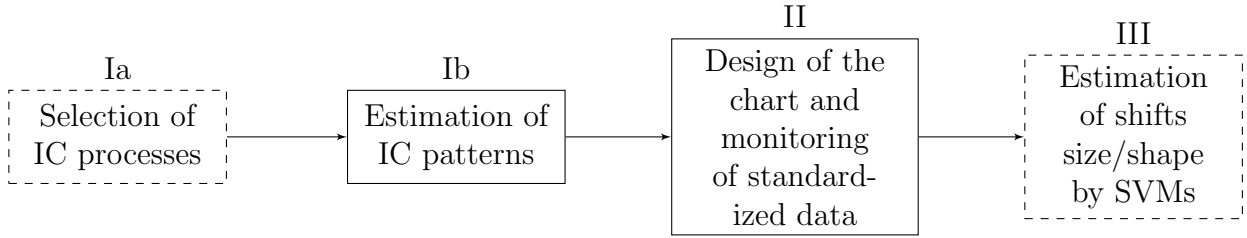 
\end{centering}


\subsection{Phase I: Estimation of the IC longitudinal patterns}
In this phase I, we automatically construct a subset of IC stations from the panel and estimate the IC patterns of the data. 

\subsubsection{Phase Ia: Selection of the IC processes}
\label{sec:Ia}

In a first stage, we need a subset (pool) of stations whose observations follow closely the median signal $M_t$. In order to find them, we calculate a stability criterion on each station.
This criterion is based on a robust version of the mean squared error (MSE) of $\hat \mu_2$:
\begin{equation}
\label{E:msesun}
STB(i) = \underset{1 \leq t\leq T}{\text{med}}[\hat \mu_2(i,t)]^2 + \underset{1 \leq t\leq T}{\text{iqr}} \hat \mu_2(i,t),
\end{equation}
where $\underset{1 \leq t\leq T}{\text{iqr}} \hat \mu_2(i,t)$ and $\underset{1 \leq t\leq T}{\text{med}} \hat \mu_2(i,t)$ denote respectively the interquartile range (IQR) and the median of the $\hat \mu_2(i,t)$ over the time for a given station $i$. 
Using these values, we can then cluster the stations and choose the cluster with the lowest values to form what we call the pool of IC stations. For this purpose, we use the $k$-means clustering ~\citep{Lloyd1957, MacQueen1967} with two clusters. \\
The pool contains deviations that will be called \emph{disparities} in the following, to be distinguished from the deviations that are supposed to be actually detected by the method. The \emph{disparities} are expected to be typically smaller and less frequent than the deviations occurring in the stations not comprised in the pool. They should be included in the design of the chart otherwise the scheme would be over-sensitive. 

The pool suffers in addition from deviations that are of similar magnitude as those of the out-of-control (OC) processes. To cope with this and preserve the detection power of our scheme, we also apply a Shewhart chart~\citep{Shewhart1931} with \emph{adaptive} confidence intervals on the data. We remove the IC observations that do not fall into one standard deviation around the cross-sectional mean ($\frac{1}{N}\sum_{i=1}^N \hat \mu_2(i,t)$). Note that we also test the method with two standard deviations instead of one and the results were similar.
We emphasize that this adaptive Shewhart would not be a substitute for our control scheme: it only removes the largest deviations at each time without taking into account the history of the observations. Therefore, contrarily to our method, it cannot detect the small and persistent shifts.

\subsubsection{Phase Ib: Estimation of the mean and the variance of the IC series}
\label{sec:Ib}
We denote by $\mu_0(t)$ and $\sigma_0^2(t)$ respectively the mean and the variance of the $\hat \mu_2 $ of the pool.
Those are estimated by the empirical mean and variance using nearest neighbours (K-NN) regression method:
\begin{equation}
\label{E:musigma}
\begin{split}
 \hat \mu_0(t)= & \frac{1}{\Delta(t)} \sum_{t'=t-\Delta(t) /2}^{t+\Delta(t) /2} \frac{1}{N_{IC}} \sum_{i_{ic}=1}^{N_{IC}} \hat \mu_2 (i_{ic}, t')  \quad s.t. \quad K = \Delta(t)  N_{IC}\\
 \hat \sigma_0^2(t)= & \frac{1}{\Delta(t)} \sum_{t'=t-\Delta(t) /2}^{t+\Delta(t) /2} \frac{1}{N_{IC}} \sum_{i_{ic}=1}^{N_{IC}} \left(\hat \mu_2 (i_{ic},t') - \hat \mu_0(t) \right)^2   \quad s.t. \quad K = \Delta(t)  N_{IC}, 
\end{split}
\end{equation}
where $i_{ic}$ denotes the index of a station of the pool. 
With K-NN regression, the temporal window $\Delta(t)$ can be adjusted to compensate the missing values of the stations, such that $ \hat \mu_0(t)$ and $\hat \sigma_0^2(t)$ are always computed on the same number ($K$) of observations. 
For appropriate use in the CUSUM chart statistics (to be defined in (\ref{E:CUSUM}) below), the data must be standardized by the IC mean and variance (as in (\ref{E:residuals}) below). Hence the number of nearest neighbors $K$ is selected to obtain the ``best'' standardization of the complete panel, in the sense that their empirical mean becomes close to zero and their empirical variance close to one.
Then, $\Delta(t)$ is chosen in time direction such that $K = \Delta(t) N_{IC}$. 
 
\subsection{Phase II: Monitoring}
\label{sec:II}

We now turn our attention to monitoring the entire panel. As a reminder, we are analyzing long-term biases denoted by $\hat \mu_2$. Using the IC mean and standard deviation $\hat \mu_0(t)$ and $ \hat \sigma_0(t)$, we standardize the (IC and OC) stations to be able to use common monitoring criteria:
\begin{equation}
\label{E:residuals}
\hat \epsilon_{\hat \mu_2}(i,t) =\frac{\hat \mu_2(i,t) -\hat \mu_0(t)} {\hat \sigma_0(t)}.
\end{equation}
Let us now focus on one station (drop the index $i$). We would like to detect indications of patterns which may relate to problems at the station. This includes persistent or gradual deviations (shifts or trends) and oscillating patterns as they may occur when the observatory is used by a rotating pool of observers each of whom has their own particular way of working. A method for accumulating small and gradual deviations is to aggregate them over time. A well known method for doing so in the context of statistical process control is the cumulative sum (CUSUM) chart~\citep{Page1961}.
The two-sided CUSUM chart applied on the residuals writes as:
\begin{equation}
\label{E:CUSUM}
\begin{split}
C_j^+= max(0,C_{j-1}^+ + \hat \epsilon_{\hat \mu_2}(t)-k) \\
C_j^-= min(0,C_{j-1}^- + \hat \epsilon_{\hat \mu_2}(t)+k),
\end{split}
\end{equation}
where  $j \geq 1$, $C_0^+ =C_0^-=0$ and $k>0$ is the allowance parameter \citep{Qiu2013}.\\
This chart gives an alert if  $C_j^+> h^+$ or $C_j^-< h^-$, where $h^-$ and $h^+$ are the control limits of the chart. Since the distribution of the residuals is almost symmetric, we use $h=h^+=-h^-$.  \\

High deviations may affect the series. Those lead to high values of the CUSUM statistics which may stay in alert for longer periods than the actual durations of the shifts. Therefore, in case of too high (resp. too low) values, we set the chart to a maximal value $2h$ (resp. $-2h$). Hence $|C^+_j|,|C^-_j| \leq 2h$.

\subsubsection{Design of the chart}
\label{sec:block-bootstrap}

As it is clear from the nature of the data, the series to be monitored have a considerable degree of autocorrelation even when they are in control. We therefore need a method for determining the control limit of the chart that takes autocorrelation into account. The block bootstrap (BB) method does this. It is based on constructing a bootstrap reference distribution by resampling blocks of data and thereby preserving the autocorrelation of the series. \\
The control limit ($h$) is adjusted here by a searching algorithm until a pre-specified rate of false positives (evaluated using the IC average run length, $ARL_0$) is reached with the desired accuracy. 
The algorithm is explained in details in Appendix \ref{app:design} (in the supplementary material) and works as follows. 
A target shift size, $\delta_{tgt}$, is first selected (it can also be estimated from the OC series as explained in Appendix \ref{app:shift_size}).
The allowance parameter is specified to $k=\delta_{tgt}/2$. 
Then, the actual $ARL_0$ is evaluated for an initial value of the control limit on IC data that are sampled from the pool by the BB procedure. 
If the actual $ARL_0$ is inferior (resp. superior) to the pre-specified $ARL_0$, the control limit of the chart is then increased (resp. decreased). This algorithm is iterated until the actual $ARL_0$ reaches the pre-specified $ARL_0$ at the desired accuracy. \\

As theoretically demonstrated in~\cite{Lahiri1999}, BB methods using non-overlapping blocks and random block lengths are more variable than those based on overlapping blocks and constant lengths. Therefore, we select the popular moving BB (MBB)~\citep{Kunsch1989, Liu1992} to obtain the best performances. \\
Since the BB preserves the serial correlation of the data inside the blocks, the length of the blocks should be selected appropriately. 
Large blocks usually model the autocorrelation of the data properly but at the same time do not represent well the variance and the mean of the series. And conversely. Using the method described in Appendix \ref{app:block_length}, the block length may be selected as the first value such that the MSE of the empirical autocorrelation of the $\hat \mu_2$ becomes stable (it corresponds to the ``knee'' (or elbow) \citep{Satopaa2011} of the MSE curve). 
This value intuitively corresponds to the smallest length which is able to represent the main part of the autocorrelation of the series. \\

The data also contain missing values. 
Among them, there are mainly large gaps since the smoothing processes in (\ref{E:e2}) removes the shortest gaps of the series. As the observing conditions could be different after a large amount of missing values (different weather conditions or instruments), we restart the scheme after each gap ($C_j^+ =C_j^-=0$). 
Blocks composed only of missing values are not used to design the chart. This may happen when some stations contain very few observations on the period studied here, either because they are ancient and stopped observing at the beginning of the period or because they have started observing only recently. 

\subsection{Phase III: Estimation of the sizes and shapes of the shifts using SVMs}
\label{sec:III}

The CUSUM gives an alert when a deviation is detected in the data but does not provide information about the characteristics (shape and size) of the shift. Such information is however valuable to assign possible causes to the shift or to adapt the type of alerts that is sent back to the observers. To that end, ~\cite{Cheng2011} appended a support vector regression (SVR) to the CUSUM to predict the magnitude of the shift after each alert. Their method is however designed for independent and identically normally distributed data that only experience jumps.
Therefore, we extend~\cite{Cheng2011} and design a method that is effective to detect the sizes \emph{and} the shapes of the deviations in the sunspot number data. This is achieved by a SVM classifier (SVC)~\citep{Burges1998} in addition to a SVR on top of the chart. 

\subsubsection{Input vector}
\label{sec:input-vector}

When an alert is triggered, the $m$ most recent observations of the stations are fed into the SVR and SVC which then predict the size and the shape of the deviation at the origin of the alert, as explained in the next subsection. 
In particular, the SVR prediction model writes as:
\begin{equation}
\label{E:SVR}
\hat \delta=f(V_{t'})=f(\hat \epsilon_{\hat \mu_2}(t'-m+1),\hat \epsilon_{\hat \mu_2}(t'-m+2),...,\hat \epsilon_{\hat \mu_2}(t')),
\end{equation}
where $t'$ denotes the time of the alert and $V_{t'}$ represents the input vector, i.e. a sequence containing the last $m$ observations of the series.  

The length $m$ of the input vector should thus be sufficiently large to contain the starting point of most of the deviations while maintaining the computing efficiency of the method. Large shifts are often quickly detected by the chart (short OC run length) while the smallest shifts may be identified only after a certain amount of time (long OC run length). Therefore, the latter require larger input vectors than the former. 
Here, $m$ is selected as an upper quantile of the OC run length distribution for a shift size equal to $\delta_{tgt}$, as explained in Appendix \ref{app:input_vector}. Hence, $m$ should be sufficiently large to allow the identification of shift sizes that are superior or equal to $\delta_{tgt}$. \\
As the SVM procedures do not support missing values, we have to impute them. Missing observations occurring at the beginning of $V_{t'}$ are simply replaced by the first valid observation encountered, while the ``intermediate'' gaps are filled by a linear interpolation. However, when there are too many of them, the analysis makes no sense. We decide to only analyze input vectors which have at least 20\% of non-missing values. 

\subsubsection{Support vector regression}

The support vector machine (SVM)~\citep{Vapnik1998} is a supervised machine-learning procedure, here used as a robust classifier and regressor to predict the shape and size of the deviations. The method has a strong theoretical basis that takes root in the optimization theory. It is able to perform efficiently non-linear classification or regression using a kernel trick that implicitly maps the data into a high dimension where the non-linear problem becomes linear. 
We only introduce the SVR in the following, since the SVC can be expressed with a similar framework. \cite{Smola2004} may also be consulted for more detailed explanations. \\ 
We denote by \{ $\bm{x}^j, \delta^j| j=1,2,...,M$ \} the $M$ training pairs. $x \in \mathcal{R}^m$ represents a training input, i.e. a series of $m$ observations that contains a deviation and $\delta \in \mathcal{R}$ is its corresponding output, the size of the deviation.  
The SVR aims at estimating the continuous regression function relating the deviating observations to the size of the shift, $f(\bm{x})$, based on the training pairs. This function writes as:
\begin{equation}
f(\bm{x})=\bm{w}^T \phi(\bm{x})+b,
\end{equation}
where $\phi$ is the non-linear function mapping the input data into the high dimensional feature space, where the regression may be expressed into a simpler linear problem. 
The coefficients $\bm{w}$ and $b$ are then estimated during the training by solving the following optimization problem: 
\begin{equation}
\label{E:regloss}
\underset{\bm{w},b}{\text{min}} \frac{1}{2}||\bm{w}||^2 + \lambda \frac{1}{M} \sum_{j=1}^M L_{\epsilon}(\delta^j, f(\bm{x}^j)),
\end{equation}
where
\begin{equation}
\label{E:loss_function}
L_{\epsilon}(\delta^j, f(\bm{x}^j)) =  
\begin{cases}
\begin{tabular}{cl}
$|\delta^j-f(\bm{x}^j)|-\epsilon$ & $\quad |\delta^j-f(\bm{x}^j)| \geq \epsilon$   \\
$0$ & otherwise.
\end{tabular}
 \end{cases}
\end{equation}
In the objective function of (\ref{E:regloss}), the parameter $\lambda$ represents a trade-off between misclassification and regularization whereas $\epsilon$ in the loss function of (\ref{E:loss_function}) is equivalent to an approximation accuracy, i.e. errors below $\epsilon$ are neglected.
This optimization problem may be rewritten with Lagrange multipliers as a dual problem and easily solved in the input space thanks to the introduction of a kernel function $K(\bm{x^j},\bm{x^k})=\phi(\bm{x^j})\phi(\bm{x^k})$. 
This kernel function is an important hyper-parameter of the method that should be carefully selected. After testing different kernels, we choose the radial basis function, to obtain the best prediction results.

\subsubsection{Creation of the training and testing sets}
\label{sec:sets}

The training and testing sets are constructed by simulations since only a limited amount of observations are available. As the SVM procedures are supposed to predict the characteristics of the shift after an alert has been raised by the CUSUM, we generate sets of series that will be first monitored by the control scheme before reaching the SVMs. When an alert will be triggered by the CUSUM, the $m$ last values of the series will be assembled and used as an input vector for the SVMs. Hence, we create series that are initially longer than $m$. Those are generated from the IC data by BB. To ensure the efficiency and the generalization of the predictions, we then add various deviations with different sizes and shapes on top of the series. \\
{\emph{Shift sizes}} The magnitudes of the shifts, $\delta$, are first randomly sampled from two half-normal distributions~\citep{distributions} supported by $[-\infty, ...,-\delta_{tgt}]$ and $[\delta_{tgt},...,\infty]$ respectively. We select the scale parameter of the half-normals equal to 3.5, a value that is sufficiently high to reproduce the highest values/deviations observed in the data. \\
{\emph{Shift shapes}:} For each $\delta$, a series $x_{ic}$ of length $T'$ is generated from the IC pool by the BB. Here, we choose $T'=500$. Three types of general deviations are then artificially constructed on top of the series: 
\begin{enumerate}
\setlength{\itemsep}{0.1cm}
\item jumps: $x(t) = x_{ic}(t)+\delta$ ;
\item drifts with varying power-law functions: $x(t) = x_{ic}(t)+\frac{\delta}{T'}(t)^{a}$, where $a$ is randomly selected in the range $[1.5, 2]$ ;
\item oscillating shifts with different frequencies: $x(t) = x_{ic}(t) \sin{(\eta \pi t)} \delta$, where $\eta$ is randomly selected in the range $[\frac{\pi}{m}, \frac{3\pi}{m}]$.
\end{enumerate}
These deviations are not data-specific to guarantee the generalization of the predictions of the characteristics of the shifts. \\
{\emph{Time of the shift}:} In the data, the shifts may happen not immediately but after an initial IC period. Therefore, we also start the monitoring after a random delay in the range $[m, 3m/2]$, to train the methods at identifying shifts appearing anywhere within the input vector. Note that the SVMs as well as the control chart should be started after $m$ observations are gathered. \\

The SVR is trained on these constructed sets to predict the size of the deviations in the continuous range $[-\infty, ...,-\delta_{tgt}, \delta_{tgt},...,\infty]$. In practice, we observe that the SVR generalizes well and can make predictions on $\mathbb{R}$ even if it was only trained in a smaller range of interest. The SVC also learns on the same sets to identify three different shapes: jumps, drifts and oscillating shifts. If a wide range of deviations are simulated, only three classes are therefore involved in the classification problem.

\section{Monitoring the composite sunspot index Nc}
\label{sec:results}

In this section, we use our methodology to solve the monitoring problem of the sunspot numbers. We do so for the composite $N_c = N_s + 10N_g$ (the same approach also works for the two components, $N_s$ and $N_g$ and is presented in the supplemental material).
We first study the low-frequency deviations on data that have been smoothed on a year to extract and analyze low-frequency patterns such as trends and persistent shifts. Then, we examine on data that have been smoothed on 27 days (one solar rotation) the higher frequency patterns such as sudden jumps. The section ends with an example of a multi-scale monitoring applied to a particular observatory. 

\subsection{Lower frequency monitoring}
\label{sec:drifts}

In the first step of low-frequency monitoring of $N_c$, we smooth the long-term bias ($\hat \mu_2$) with a window length of one year as described in Section~\ref{sec:bias}. As explained in Section~\ref{sec:Ia}, the network of stations is first reduced to a pool of 119 in-control (IC) stations. 
In the next step,  we extract the IC mean and standard deviation using the K-NN regression described in Section~\ref{sec:Ib}. The selection mechanism finds $K=4600$ for this step. The resulting mean and standard deviation are then used to standardize all series. 

In the second stage, we use the block bootstrap method described in Section~\ref{sec:block-bootstrap} to calibrate the CUSUM chart at an average run length of 200. This requires choosing the block length first. In our situation, a choice of two solar rotations (54) appears appropriate. It is longer than the lifetime of most sunspots but not too long for practical use. The calibration then leads to a control limit of $h=19$ and a target shift size of $\delta_{tgt}=1.5$. \\

\begin{figure}[!htb]
	\centering
	\begin{subfigure}{0.48\textwidth}
		\centering
		\includegraphics[scale=0.48]{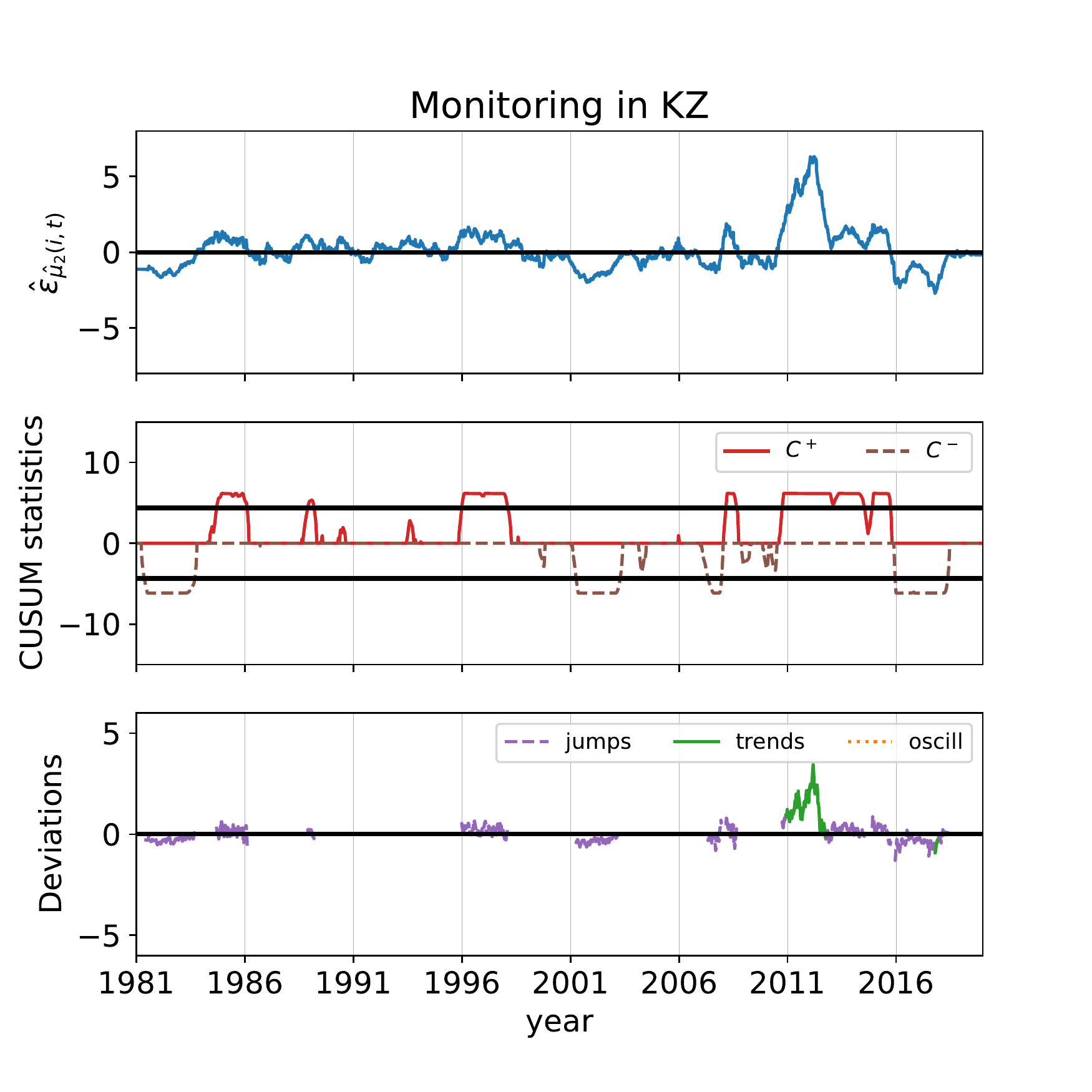}
		\caption{}
		\label{fig:drifts_KZ}
	\end{subfigure}
	\begin{subfigure}{0.48\textwidth}
		\centering
		\includegraphics[scale=0.48]{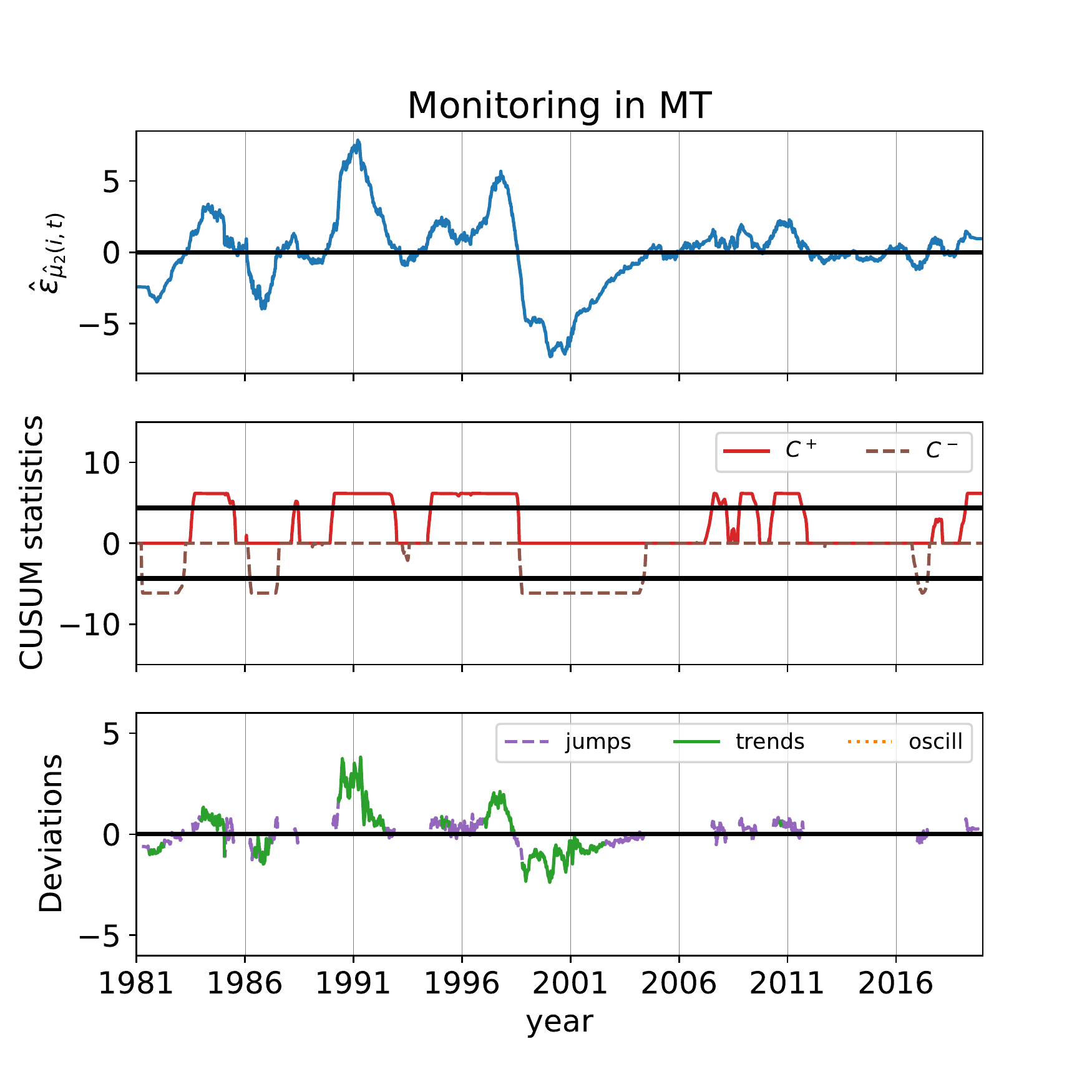}
		\caption{}
		\label{fig:drifts_MT}
	\end{subfigure}
\caption{\footnotesize{ (a) Upper panel: the residuals $\hat \epsilon_{\hat \mu_2(i,t)}$ for $N_c$ smoothed on one year from the KZ over the period studied (1981-2019). In addition to their disparities, the residuals also contain the actual deviations of the station, which have been removed for the design of the chart as explained in Section~\ref{sec:Ia}. Middle panel: the (two-sided) CUSUM chart statistics applied on the residuals in \emph{square-root scale}. The control limits of the chart are represented by the two horizontal thick lines. Lower panel: the characteristics of the deviations predicted by the SVR and SVC after each alert. (b) Similar figure for MT over the same period. }}
\label{fig:drifts}
\end{figure}

Finally, the support vector method for extracting and classifying out-of-control patterns is deployed.
It is composed of a SVR to predict the size of the shifts and a SVC to classify the shape of the encountered deviations. 
We obtain them by creating a set of artificial series of 500 values generated from the IC pool by the BB. These give us series as we would observe in reality including correlations. We then artificially add jumps, trends, and oscillating shifts to them as described in Section~\ref{sec:sets}. These series are then fed to the CUSUM chart which identifies the out-of-control observations. When an alert is triggered by the chart, an input vector containing the $m$ last observations of the series is assembled. This input vector is then analyzed by the SVR and SVC for predicting the characteristics of the shift. In our case, we harvested 63000 such series from the IC pool and we enriched them with artificial patterns. We then calibrated SVR and SVC models by splitting this set into a training set (80\%) and a testing set (20\%). 
 
The length of the input vector is specified here at $m=80$, as explained in Section~\ref{sec:input-vector}. This value of 80 corresponds to the 90-th quantile of the OC run length distribution, for a shift of size $\delta_{tgt}$.
The other parameters of the support vector machines are automatically selected from a searching interval to obtain the best prediction results. Those are evaluated using the mean absolute percentage error (MAPE) for the regression and the accuracy for the classification problem, see Appendix~\ref{app:performances}. With this method, the regularization parameter $\lambda$ of the classifier and regressor is set to 13 and the accuracy error $\epsilon$ of the SVR is fixed at 0.001.  
The performances of the SVMS are presented in Appendix~\ref{app:perfo_drifts}.
Overall, they are sufficient to achieve our goals: identify the origins of the deviations. \\

Figure~\ref{fig:drifts} shows results for two stations labeled KZ\footnote{The Kanzelh{\"o}he Observatory in Austria.} and MT\footnote{The National Observatory of Japan, in Mitaka (Tokyo)}. The observatory of KZ is rather stable and belongs to the IC pool. It relies on a stable team of well trained observers. The observatory MT is generally less stable and shows a severe downward trend around 1998. 
The cause of this downward trend could be traced to the replacement of visual counts from the direct optical solar image by automatic computerized counts based on digital images from a CCD camera. Given the image sensor technology then available, the spatial resolution of the images was limited, and many small spots that were fully detected in earlier visual observations were not detected anymore by the new equipment.

\FloatBarrier
\subsection{Higher frequency monitoring}
\label{sec:jumps}

The method described above can also be applied to the biases ($\hat \mu_2$) smoothed on a shorter time window such as the duration of a solar cycle (27 days). 
Here the selection of the IC pool yields 100 stations and the number of nearest neighbors comes out to $K=2400$. This number is smaller than before since we are working at a higher frequency.
The block bootstrap and SVMs with the same settings as above can be used to calibrate the CUSUM chart. 
For $\delta_{tgt}=1.4$, the control limit of the chart is selected at $h=13$ to obtain an average run length of 200. 
The length of the input vector is fixed here at $m=70$, which corresponds to the 90-th quantile of the OC run length distribution. \\


\begin{figure}[!htb]
	\centering
	\begin{subfigure}{0.48\textwidth}
		\centering
		\includegraphics[scale=0.48]{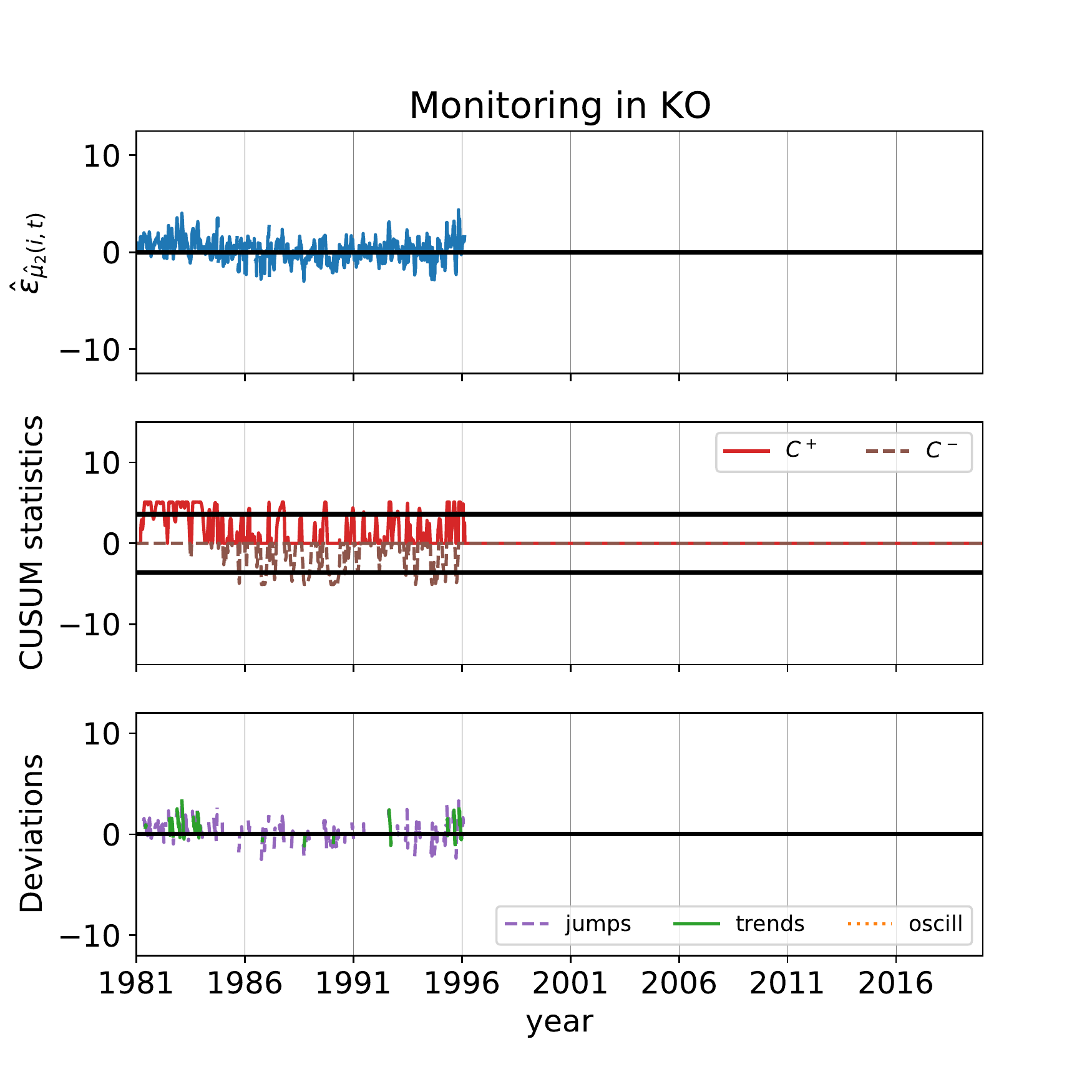}
		\caption{}
		\label{fig:jumps_KO}
	\end{subfigure}
	\begin{subfigure}{0.48\textwidth}
		\centering
		\includegraphics[scale=0.48]{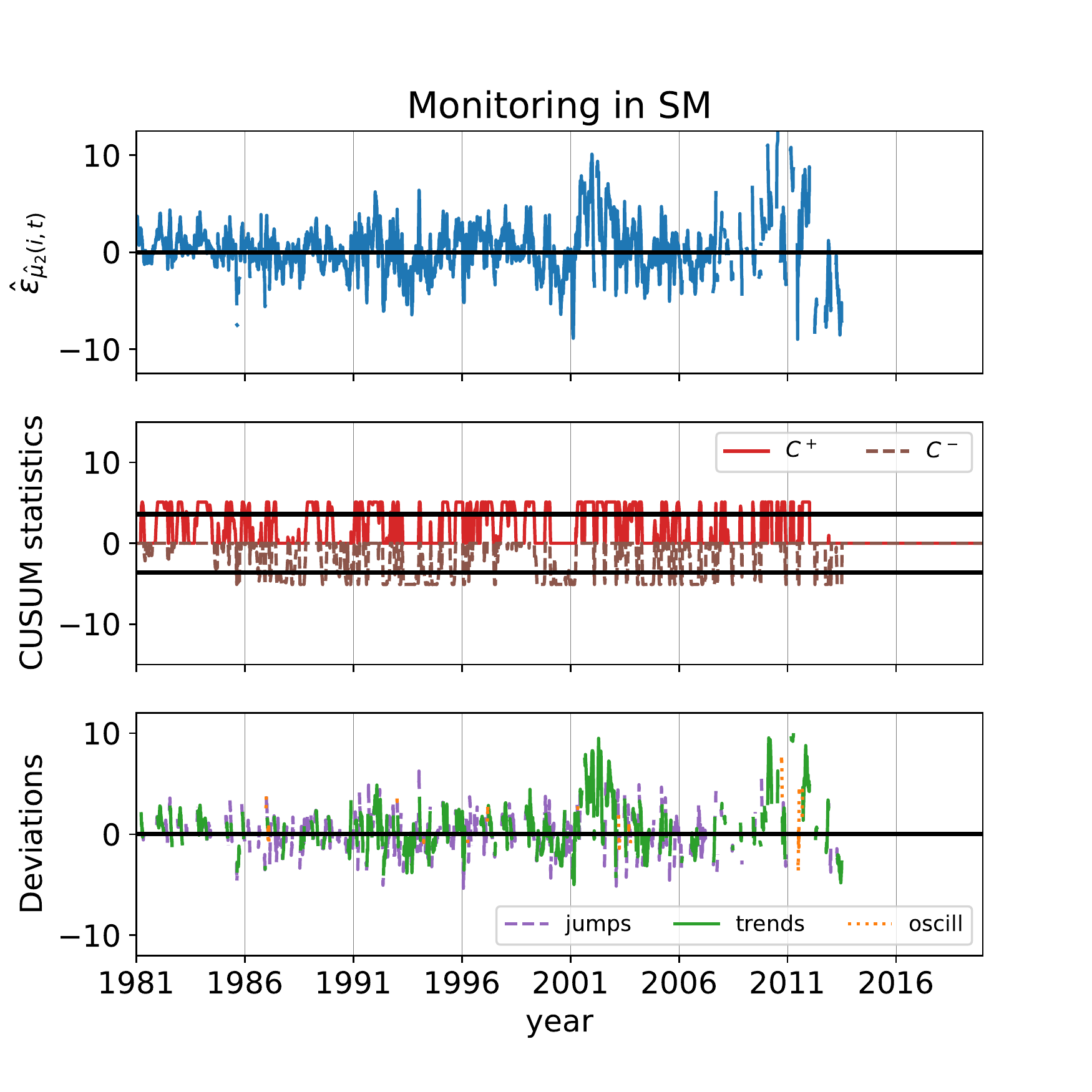}
		\caption{}
		\label{fig:jumps_SM}
	\end{subfigure}
\caption{\footnotesize{ (a) Upper panel: the residuals $\hat \epsilon_{\hat \mu_2(i,t)}$ for $N_c$ smoothed on 27 days for KO over the period studied (1981-2019). In addition to their disparities, the residuals also contain the actual deviations of the station, which have been removed for the design of the chart as explained in Section~\ref{sec:Ia}. Middle panel: the (two-sided) CUSUM chart statistics applied on the residuals in \emph{square-root scale}. The control limits of the chart are represented by the two horizontal thick lines. Lower panel: the characteristics of the deviations predicted by the SVR and SVC after each alert. (b) Similar figure for SM over the same period. }}
\label{fig:jumps}
\end{figure}

Figure \ref{fig:jumps} shows the methodology applied to KO\footnote{Name of the observer known to the authors, kept for privacy.} and SM\footnote{The observatory of San Miguel in Argentina.}. KO was a Japanese observatory run by a single dedicated observer whose records (which stopped during 1996) were very stable.   
On the contrary, SM is a severely OC station 
that experiences large known deviations \citep[Figure~12]{Mathieu2019}.
 The large variations observed in SM are likely caused by the rotation of several observers involved in the counting process. In some countries, the public observatories have also an educational function. Their team of regular observers are usually small and are often completed by student or amateur astronomers that are frequently replaced, which causes large variations. 
Unfortunately, we could not identify more precisely the origin of the deviations since their observations stopped several years ago and we have not succeeded in contacting them yet. The lack of information is a common problem we face when investigating past deviations in stations that are now inactive and is therefore worth mentioning. \\ 
As we see in the figure, the biases vary a lot at 27 days, a scale which is close to the short-term regime. The actual monitoring should be based on a larger scale otherwise some stations such as SM would receive almost constant alerts. If a particular deviation is detected at higher scale (such as one year), it might be interesting however to analyze it at 27 days, to better identify its origin. 

\FloatBarrier
\subsection{Multi-scale monitoring}
\label{sec:discussions}

\begin{figure}[!htb]
	\centering
	\begin{subfigure}{0.48\textwidth}
		\centering
		\includegraphics[scale=0.48]{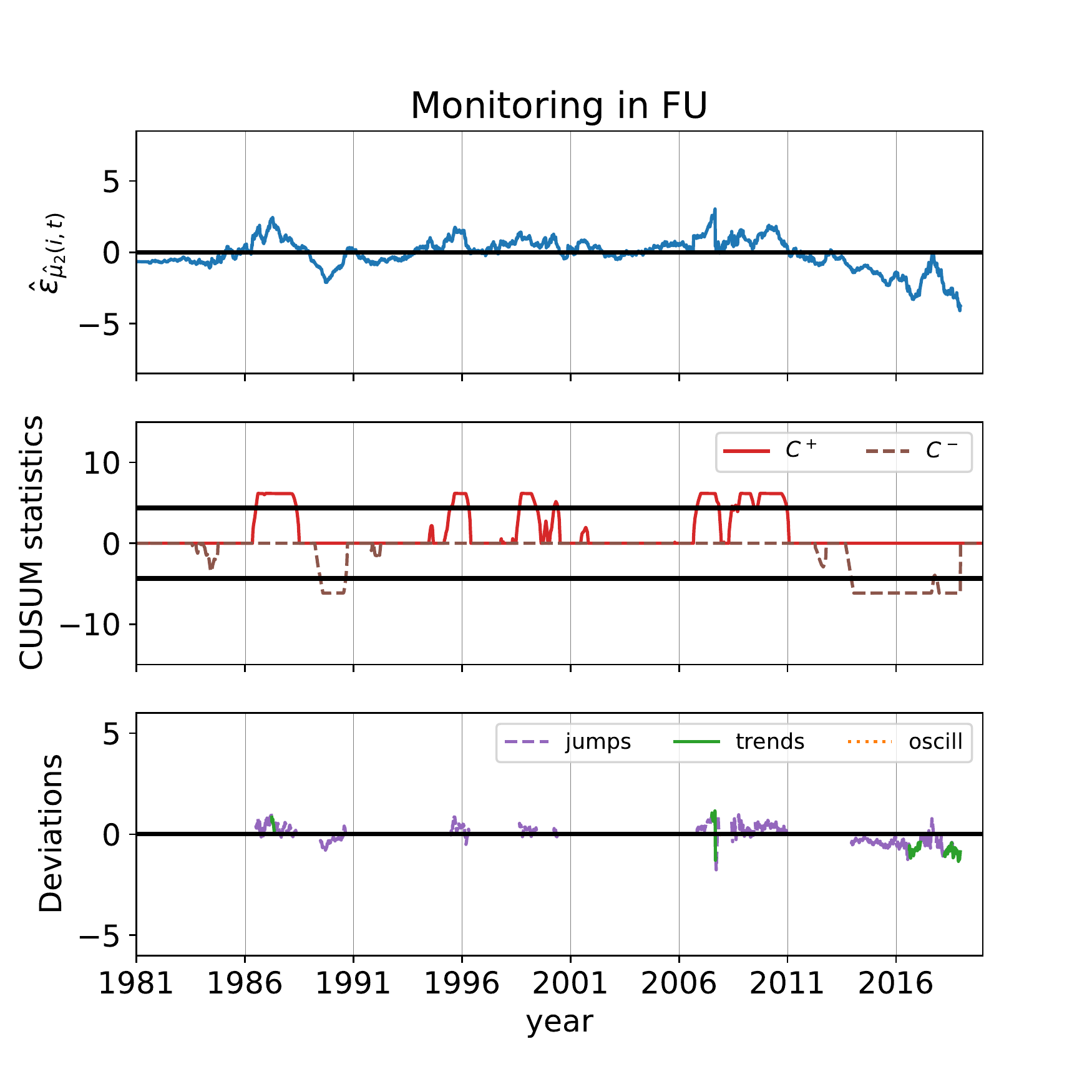}
		\caption{}
		\label{fig:drifts_FU}
	\end{subfigure}
	\begin{subfigure}{0.48\textwidth}
		\centering
		\includegraphics[scale=0.48]{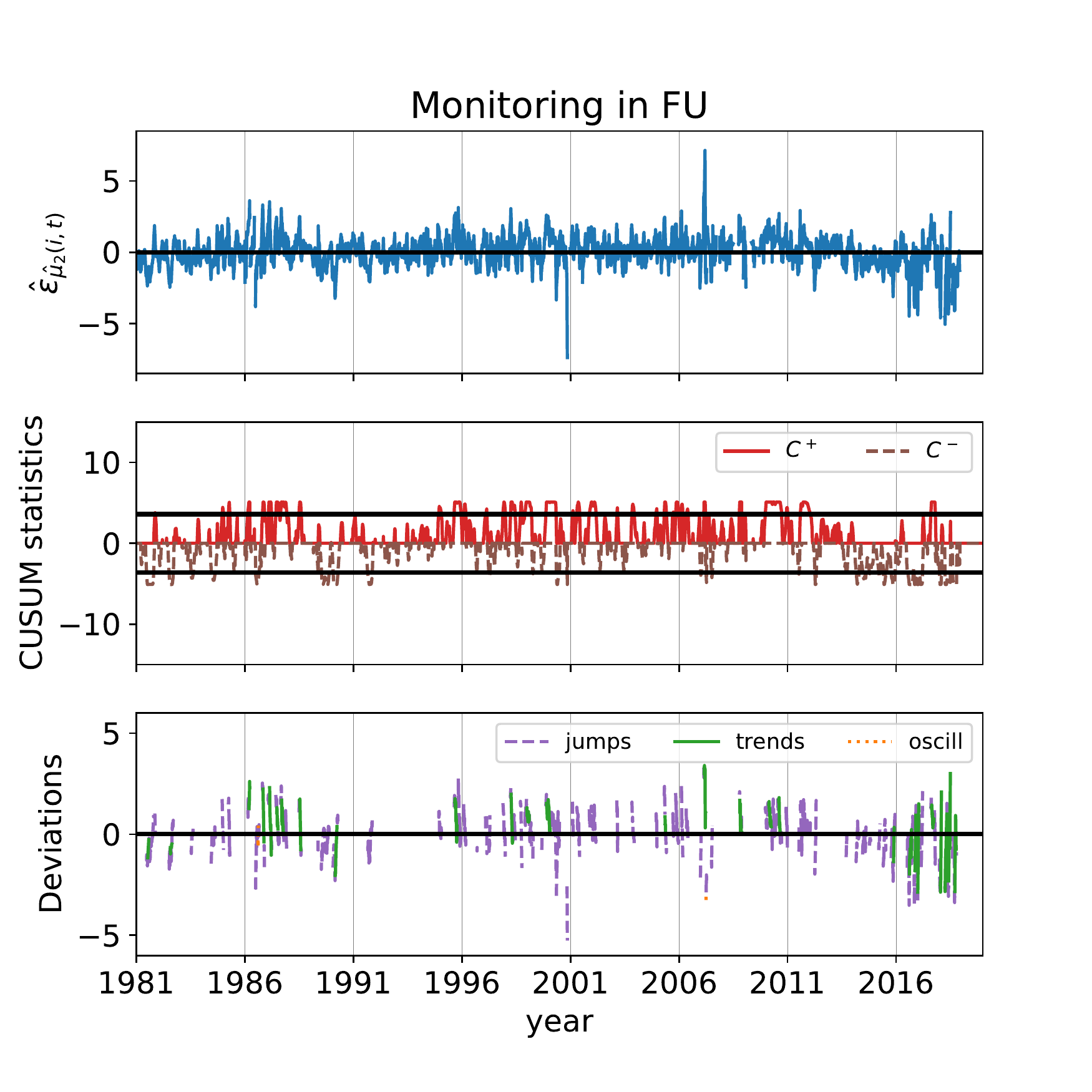}
		\caption{}
		\label{fig:jumps_FU}
	\end{subfigure}
\caption{\footnotesize{ (a) Upper panel: the residuals $\hat \epsilon_{\hat \mu_2(i,t)}$ for $N_c$ smoothed on 365 days from FU over the period studied (1981-2019). In addition to their disparities, the residuals also contain the actual deviations of the station, which have been removed for the design of the chart as explained in Section~\ref{sec:Ia}. Middle panel: the (two-sided) CUSUM chart statistics applied on the residuals in \emph{square-root scale}. The control limits of the chart are represented by the two horizontal thick lines. Lower panel: the characteristics of the deviations predicted by the SVR and SVC after each alert. (b) Similar figure for the values of $\hat \epsilon_{\hat \mu_2(i,t)}$ smoothed on 27 days in FU. }}
\label{fig:FU}
\end{figure}

\begin{figure}[hbt]
	\centering
		\includegraphics[scale=0.4]{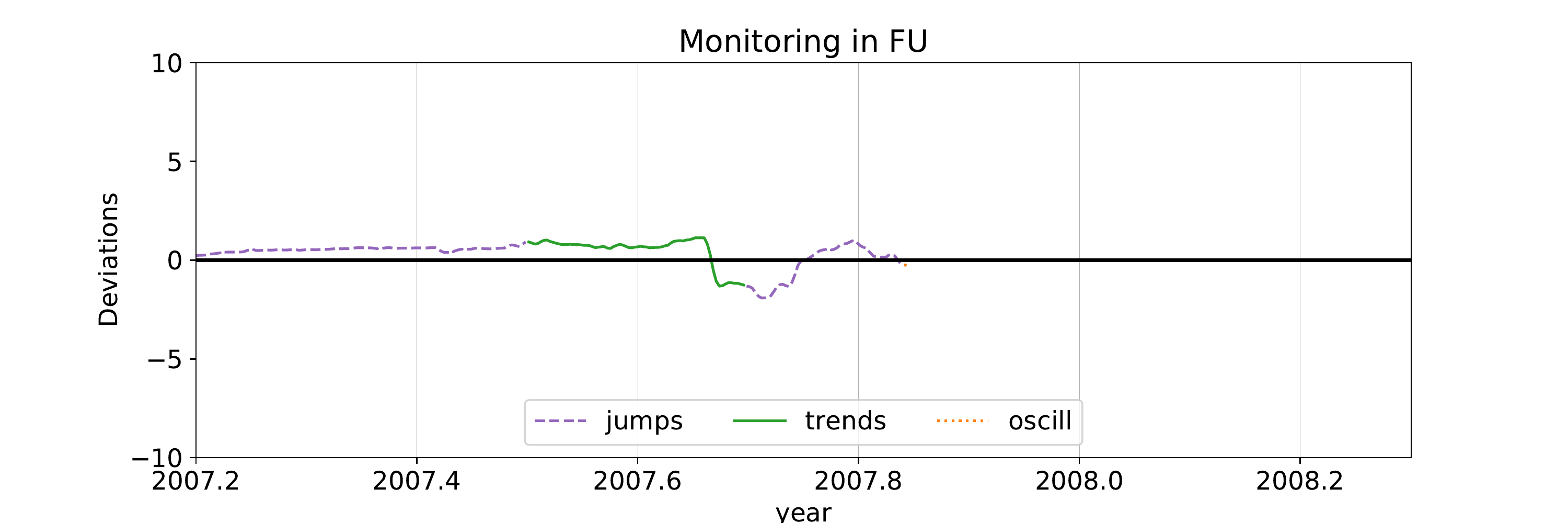}
		\caption{\footnotesize{Sizes and shapes of the deviations (taken from Figure \ref{fig:FU}a) predicted by SVMs in FU over 2007-2008.}}
	\label{fig:zoom}
\end{figure}

Figures \ref{fig:drifts}  and \ref{fig:jumps} display instances of a stable IC station included in the pool and a typical out-of-control observatory for the high- and low- frequency monitoring respectively.
To better grasp the motivations of a multi-scale monitoring, the method is applied to the data smoothed on 27 days and one year of FU\footnote{Name of the observer known to the authors, kept for privacy.} in Figure \ref{fig:FU}. 
The FU station is composed of a single dedicated observer in Japan, who has observed without interruption since 1968 until today, producing one of the longest individual series. His observations are included in the IC pool but yet suffer from recent deviations. In particular, the upward deviation (which looks like a spike) reported in 2007 in FU \citep{slides_frederic} as well as the downward drift occurring after 2014 are well identified in Figure~\ref{fig:drifts_FU}. Figure \ref{fig:zoom} shows a zoom of Figure~\ref{fig:drifts_FU} on the time period from 2007 to 2008.
After a progressive upward shift, the station experiences a rapid downward trend over five days. This trend, which looks like a jump in the whole period view, it thus correctly classified by the SVC. 
Even by taking a closer look on the figure however, it remains difficult to precisely identify the origin of the shifts on data that are smoothed on a year. 
By looking at a smaller scale of 27 days in Figure~\ref{fig:jumps_FU}, we can better characterize the shift in 2007 as a short event and pinpoint its location. 
After investigations, this deviation appears to be related to a small over-count that appeared in early 2007 (three groups were reported in FU while most of the network only observed two groups) while the drift might be associated to the health condition of the observer.
Note that the long-term biases are not defined (i.e. set to missing values) when the median of the network is equal to zero, see (\ref{E:e2}). This regime corresponds to those of the variability at minima, represented by $\epsilon_3$. Due to the smoothing procedure of (\ref{E:e2}), the deviations that appeared close to solar minima, such as the jump in FU, are thus particularly visible. \\
As shown in the figures, the monitoring and the SVM procedures can cope with a large variety of shifts ranging from small and persistent deviations to large oscillating shifts. The procedures automatically detect major deviations recently discovered by hand as mentioned above. More identified prominent deviations as well as results for other stations are shown in Appendix \ref{app:figures}. 
In addition, the chart also unravels many other shifts, typically smaller, that are otherwise difficult to identify.

Note that the figures represent past observations, which have not been monitored by any control scheme. Consequently, the stations may stay in alert for long consecutive periods. 
If this method is applied on future observations, any major deviation will be promptly corrected. 
Therefore we expect better results for data that have already been monitored.

\section{Conclusion and perspectives}
\label{sec:conclusion}

We presented a nonparametric control scheme to monitor challenging and important datasets related to the observations of sunspots across a wide network of stations. The approach allows us to deal with the missing values, autocorrelations, and non-normality of the data, to detect and classify station-related anomalies on different time scales. The procedure is based on a particular choice of methods for smoothing, robust anomaly detection, and anomaly classification. Other methods exist for these steps; yet we believe that our choices are particularly suitable for the problem at hand.

The features of our approach are multi-scale smoothing, CUSUM charting, SVM classification and detailed graphical displays. They also include an automatic pre-selection of an in-control pool and the powerful calibration of the chart using block bootstrap procedures.  The associated advantages are robustness, flexibility, automation, and guided interpretation of results. 
The method allows us to detect and identify the causes of major deviations that occurred in the series.
We have seen that monitoring on at least two-time scales is essential to capture these anomalies. Some patterns first attract interest on a long-term scale but it is at the short-term scale that their potential root-causes can be suggested.
The method also identifies a wide range of deviations unseen in previous analyses. Most of them have not been related to specific causes yet but will soon be investigated. \\

This automated method allows us and the researchers at the Royal Observatory who are in charge of producing the International Sunspot Number
to have a harmonized view across the network of stations. It provides a way to give specific and targeted advice to the observers. 
As demonstrated in this paper, the method also delivers easy to interpret graphical displays which facilitate root cause analysis of deviations.
The complete re-examination of past data of the whole panel has just started. When they will be finished, these analyses will allow us to arrive at a cleaner data stream and to release an improved version of the International Sunspot Number. 
Additionally, the implementation of the method in the continuous surveillance of future observations will lead to a faster detection and identification of inconsistencies, their elimination by better observer training or equipment maintenance, and finally to a more precise determination of the sunspot numbers in the future. \\

Our methodology has the potential to adapt to other type of panel data such as those observed in the manufacturing or financial industry, which remains to be investigated.

\FloatBarrier
\bigskip
\begin{center}
{\large\bf SUPPLEMENTARY MATERIAL}
\end{center}

\begin{description}

\item[Python package (codes)] The subset of data and the codes that we used in this paper are available at \url{https://github.com/sophiano/SunSpot}.

\item[Algorithms] The pseudo-algorithms to design the CUSUM chart, to select the target shift size, to choose the block length and the length of the input vector are explained in the supplementary material in Appendix A. 

\item[Performance criteria] The performances of the support vector procedures for the high and low frequency monitoring are displayed in the supplementary material in Appendix B. 

\item[Additional figures] Additional analyzes and figures are also provided in the supplementary material in Appendix C. 

\end{description}

\section{Acknowledgments}
\label{sec:thanks}
The first author gratefully acknowledges funding from the Belgian Federal Science Policy Office (BELSPO) through the BRAIN VAL-U-SUN project (BR/165/A3/VAL-U-SUN). S. Mathieu, L.Lef{\`e}vre and F.Clette also acknowledge financial support from the International Space Science Institute (ISSI, Bern, Switzerland) via the International Team 417 ``Recalibration of the Sunspot Number Series'', chaired by M. Owens and F. Clette\footnote{\url{https://www.issibern.ch/teams/sunspotnoser/}}. Computational resources have been provided by the Consortium des {\'E}quipements de Calcul Intensif (CECI), funded by the Fonds de la Recherche Scientifique de Belgique (F.R.S.-FNRS) under Grant No. 2.5020.11 and by the Walloon Region.
The authors also thank Hisashi Hayakawa and all observers that provided their help in the search for the origins of the deviations. 

\bibliographystyle{apalike}
\bibliography{thesis_biblio}

\end{document}


\def\spacingset#1{\renewcommand{\baselinestretch}%
{#1}\small\normalsize} \spacingset{1}

\spacingset{1.5} 

\FloatBarrier
\bigskip
\begin{center}
{\large\bf SUPPLEMENTARY MATERIAL}
\end{center}

This supplementary material provides some more details about the monitoring scheme as well as more examples of monitored stations.

%
%
%
%
%


\appendix
\section{Algorithms}
\label{app:Algorithms}

\subsection{Design of the CUSUM chart}
\label{app:design}

The performances of the chart are usually measured using the concept of the average run length (ARL). 
The in-control (IC) ARL, denoted by $ARL_0$, is the mean value of the number of samples collected from the beginning of the process to the occurrence of a false alert. It represents the rate of false positives of the chart (it is similar to the concept of type I error in hypothesis testing context). 
The out-of-control (OC) ARL, denoted $ARL_1$, corresponds to the mean value of the number of samples collected from the appearance of a shift to the alert of the chart. It embodies the detection power of the chart (similar to the concept of type II error in hypothesis testing context). 
In practice, it is hardly possible to design a chart with $ARL_0$ as large as possible while maintaining at the same time a small value of $ARL_1$. Therefore, as in hypothesis testing, the parameters of the chart are tailored to reach the maximal detection power for a fixed rate of false positives. \\

Hence, the control limit of the CUSUM chart may be adjusted on the data using a searching algorithm until a pre-specified value of $ARL_0$ is reached to a desired accuracy. Since we do not have information about the distribution or the autocorrelation model of the data, the chart is designed using the block bootstrap as explained in Algorithm \ref{PA:searching}. This method is based on an algorithm of~\cite{Qiu2013}, Section 4.2.2 that is tailored to i.i.d. normal data.  The procedure works as follows.
The allowance parameter, $k$, is first specified in advance to $k=\delta_{tgt}/2$~\citep{Moustakides1986},  where $\delta_{tgt}$ is the target shift size. From initial value of the control limit, the actual $ARL_0$ is then computed on B series that are sampled with repetition from the pool of $N_{IC}$ standardized IC processes. If the actual $ARL_0$ is inferior (resp. superior) to the pre-specified $ARL_0$, the control limit of the chart is increased (resp. decreased).
This algorithm is iterated until the actual $ARL_0$ reaches the pre-specified $ARL_0$ at the desired accuracy. 

\begin{algorithm}[H]
\DontPrintSemicolon
\LinesNotNumbered 
\tcc{Adjust the control limit of the chart using the block bootstrap method}
Select values for: \;
[$h_U$,$h_L$], the interval where $h$ is searched on \;
$ARL_0$, the desired value of the in-control ARL \;
$\rho$, the accuracy to reach $ARL_0$ \;
$\delta$, the target shift size \;
\;
$k=\delta/2$ \;
\While{$|ARL-ARL_0|>\rho$}{
$h=\frac{h_U+h_L}{2}$ \;
\For{b in (1, B)}{
\tcp{sample data with repetitions from the $N_{IC}$ (IC) processes}
resample $\hat \epsilon_{\hat \mu_2}(i_{IC},t)$ per blocks \; 
compute the chart statistics $C^+$ and $C^-$ on the resampled data \;
\uIf{the chart gives an alert ($C^+>h$ or $C^-<-h$)}{
$RL[b]$ $=$ time of alert \; }
\Else{
$RL[b]$ $=$ large number \;
}
}
$ARL$ = mean($RL$) \;
$h_U=(h_L+h_U)/2$, $h_L=h_L$ if $ARL_0 > ARL$ \;
$h_L=(h_L+h_U)/2$, $h_U=h_U$ if $ARL_0 < ARL$ \;
}
$h=\frac{h_U+h_L}{2}$ \;
\;
 \caption{Searching algorithm for adjusting the control limit of the CUSUM chart}
\label{PA:searching}
\end{algorithm}
\noindent
\vspace{1.45cm}

\subsection{Selection of the target shift size}
\label{app:shift_size}

The target shift size $\delta_{tgt}$ may be estimated, without prior information, on the out-of-control (OC) series by a recursive method described in Algorithm \ref{PA:size}. This procedure works as explained below.
From an initial value of $\delta_{tgt}$, denoted by $\delta_0$, we compute the control limit of the chart for $k=\delta_0/2$ as explained in Algorithm \ref{PA:searching}. Then, the chart is applied on data that are resampled from the OC series by the block bootstrap procedure.
For each sample, the magnitude of the deviation is computed after the alert using the following formula~\citep{Montgomery2005}:
\begin{equation}
\label{E:Montgomery}
\hat \delta =    \begin{cases}
\begin{tabular}{cl}
$k+\frac{C_i^+}{N^+}$ & if $C_i^+>h$  \\
$-k-\frac{C_i^-}{N^-}$ & if  $C_i^-<-h$,
\end{tabular}
    \end{cases}
\end{equation}
where $\hat \delta$ is the estimated shift size expressed in standard deviation units and $N^+$ (resp. $N^-$) represents the number of observations where the CUSUM statistics $C_i^+$ (resp. $C_i^-$) has been non-zero. 
Although valid for uncorrelated and normally distributed observations, the above formula can still be used in general to provide a rough approximation of the shifts size (those will be estimated afterward by support vector regression). 
Then, a new value for $\delta_{tgt}$ is computed as a specified quantile of the shifts size distribution. 
The algorithm is iterated a few times until convergence. 

\begin{algorithm}[H]
\DontPrintSemicolon
\LinesNotNumbered 
\tcc{Estimate a target shift size using the block bootstrap method}
Select values for: \;
$\delta_0$, an initial value of the target shift size \;
$\rho$, the accuracy of the algorithm's convergence \; 
\;
$\delta_{prev}$ $=$ 0 ; $\delta$ $=$ $\delta_0$ \;
\While{$|\delta-\delta_{prev}|>\rho$}{
Compute the control limit $h$ using algorithm 1 for $k=\delta/2$ \;
$\delta_{prev}$ $=$ $\delta$ \;
\For{b in (1, B)}{
\tcp{sample data with repetitions from the OC processes}
resample $\hat \epsilon_{\hat \mu_2}(i_{OC},t)$ per blocks \; 
compute the chart statistics $C^+$ and $C^-$ on the resampled data \;
\uIf{the chart gives a positve alert ($C^+>h$)}{
$shift\_size[b]$ $=$ $k + \frac{C^+}{N^+}$ (Montgomery's formula)\; }
\uIf{the chart gives a negative alert ($C^-<-h$)}{
$shift\_size[b]$ $=$ $-k - \frac{C^-}{N^-}$ (Montgomery's formula)\; }
}
$\delta$ = $quantile$($shift\_size$) \;
}
$\delta$ 
\;
 \caption{Pseudo-algorithm to estimate the target shift size}
\label{PA:size}
\end{algorithm}
\noindent
\vspace{1.45cm}

\subsection{Choice of the block length}
\label{app:block_length}

The length of the block is an important parameter that depends on the distribution and the autocorrelation of the data. 
It may be selected using Algorithm \ref{PA:block_length}. This algorithm works as follows. 
For each block length tested over a specified range, the method resamples $B$ series of observations using the block bootstrap. Then, it computes the mean squared error (MSE) of the empirical mean, standard deviation and autocorrelation at different lags of the resampled series (with respect to the original data). 
Small block lengths appear to represent the variance and the mean of the data properly (the MSE of the mean and the variance increases with the block length). Reversely, large block lengths better account for the autocorrelation of the data (the MSE of the autocorrelation decreases when the block length increases). The appropriate value for the block length is finally selected as the ``knee'' (or elbow) \citep{Satopaa2011} of the curve, i.e. the first value such that the MSE of the autocorrelation becomes stable.
This value intuitively corresponds to the smallest length which is able to represent the main part of the autocorrelation of the series.

\begin{algorithm}[H]
\DontPrintSemicolon
\LinesNotNumbered 
\tcc{Estimate the length of the blocks for the block bootstrap procedure}
Select values for: \;
$start$, a starting value for the block length \;
$stop$, a stopping value for the block length \;
$step$, a step value for the block length \;
$lag\_max$, the maximal lag up to which the autocorrelation is evaluated  \;
\;
\For{$L$ in ($start$, $stop$, $step$)}{
\For{$station$ in (1, $N_{IC}$)}{
\For{$b$ in (1, $B$)}{
\tcp{sample data from the IC series 'station'}
$boot$ $=$ resample $\hat \epsilon_{\hat \mu_2}(station,t)$ per blocks of length $L$ \; 
\tcp{compute the autocorrelation of the resampled series until lag\_max}
$autocorr\_boot[b, lag]$ $=$ autocorrelation($boot$, $lag\_max$) \;
}
\tcp{compute the MSE of the autocorrelation in a station for each lag and take the mean over the lags}
$mse\_autocorr\_station[station]$ $=$ mean(MSE($autocorr\_boot$), lag) \;
}
\tcp{compute the mean of the autocorrelation over all stations}
$mse\_autocorr[L]$ $=$ mean($mse\_autocorr\_station$) \;
}
$block\_length$ $=$ knee($mse\_autocorr$)
\;
 \caption{Pseudo-algorithm to estimate an optimal value for the block length}
\label{PA:block_length}
\end{algorithm}
\noindent
\vspace{1.45cm}

\subsection{Length of the input vector}
\label{app:input_vector}

The length $m$ of the input vector may be selected as an upper quantile of the out-of-control (OC) run length distribution as described in Algorithm \ref{PA:m}. 
This method may be explained as follows. 
We first select a value for $\delta_{tgt}$, the shift size that we aim to detect.
In-control (IC) observations are then sampled by the block bootstrap procedure and shifted by a jump of size $\delta_{tgt}$. The run lengths of the chart are later computed on these artificial series. 
The length of the input vector is finally selected as an upper quantile of the run length distribution. Different quantiles are evaluated in the range [0.5, 1] and the optimal quantile is selected as the ``knee''~\citep{Satopaa2011} of the curve. 
The aim of the procedure is to choose $m$ sufficiently large to ensure that the starting point of most of the shifts of size $\delta_{tgt}$ are contained within the input vector, while maintaining the computing efficiency of the method. Hence, the SVM procedure will be able to predict efficiently the characteristics of the shifts whose sizes are equal to or larger than $\delta_{tgt}$. 

\begin{algorithm}[H]
\DontPrintSemicolon
\LinesNotNumbered 
\tcc{Estimate an optimal value for $m$, the length of the input vector }
Select value for: \;
$\delta_{tgt}$, the shift size that we aim to detect (algorithm 2 can be used) \;
\;
Compute the control limit $h$ using algorithm 1 for $k=\delta_{tgt}/2$ \;
\For{b in (1, B)}{
\tcp{sample data with repetitions from the IC processes}
data $=$ resample $\hat \epsilon_{\hat \mu_2}(i_{IC},t)$  per blocks \; 
\tcp{add a simulated shift of size $\delta_{tgt}$ on the resampled data} 
data $+$ $\delta_{tgt}$ \;
compute the chart statistics $C^+$ and $C^-$ on the resampled data \;
\uIf{the chart gives an alert ($C^+>h$ or $C^-<-h$)}{
$RL_1[b]$ $=$ time of alert \; }
\Else{
$RL_1[b]$ $=$ large number \; }
}
$m$ = knee($RL_1$) \;
\;
 \caption{Algorithm to select the size of the input vector}
\label{PA:m}
\end{algorithm}
\noindent
\vspace{\baselineskip}

\section{Performance criteria}
\label{app:performances}

The support vector machine (SVM) prediction results can be measured with different criteria. 
The SVR predictive ability may be evaluated on the testing set with the mean absolute percentage error (MAPE): 
\begin{equation}
MAPE=\frac{1}{M}\sum_{j=1}^M \big| \frac{|\delta^j| - |\hat \delta^j|}{|\delta^j|} \big| \times 100\%,
\end{equation}
where $\hat \delta^j$ is the shift size predicted by the SVR (which can be positive or negative). Small values of MAPE are desirable since they correspond to predictions close to the actual shift sizes. \\
The performance of SVC may be evaluated by the classification accuracy: 
\begin{equation}
ACCURACY=\frac{1}{M}\sum_{j=1}^M \mathbbm{1}_{\{\hat \delta_{c3}^j = \delta_{c3}^j \}} \times 100\%,
\end{equation}
where $\hat \delta_{c3}^j$ denotes the shape of the deviation (jump, drift, or oscillating shift) predicted by the SVC, and  $\delta^j_{c3}$ is the true deviation shape. A high accuracy is desired since it corresponds to a close match between the predicted and the actual shapes of the shifts. \\
The accuracy is thus a performance measure of the classifier for all shapes. 
The confusion matrix may also be computed to obtain a detailed view of the performances of the classifier, class by class. The columns of this matrix represent the prediction labels (here the predicted shapes) whereas the rows are the true labels (the true shapes). Therefore, the elements of the matrix on the main diagonal correspond to the numbers of correct classification per class, while the other entries show the number and the type of classification errors. 

\FloatBarrier
\subsection{Lower frequency monitoring}
\label{app:perfo_drifts}

The performances of the monitoring problem described in Section~\ref{sec:drifts} are presented here.
After training, the SVR shows a MAPE of around 26 and the SVC has an accuracy of around 95\% on the testing set. \cite{Cheng2011} present MAPE values for predicting the sizes of simple jumps in normally distributed data using SVR. Those are of the same magnitude that ours. Considering the various types of deviations that are used to train the SVMs, the performances of our method are acceptable. 
The confusion matrix is written in Table~\ref{tab:cm3} for the 12600 testing pairs. With a perfect classifier, the matrix would be diagonal with elements equal to 4200. As can be seen, the most frequent errors are oscillating shifts predicted as jumps. Those account for 482 testing pairs and represent 3.8\% (482/12600) of the testing set. 
These errors are expected since a series of consecutive jumps with different sizes may look similar to an oscillating shift.

\begin{table}[hb]
\begin{center}
\begingroup
\begin{tabular}{lcccc}
\cline{1-4}
\cline{1-4}
Deviation & \multicolumn{3}{c}{Predicted} & \\
\cline{2-4}
True value & jump & drift & oscillating shift & \\
\cline{1-4}
jump & 4192 & 1 & 7 & 4200\\
drift &  34 & 4166 & 0  & 4200 \\
oscillating shift & 482 & 1 & 3717 & 4200\\
\cline{1-4}
& 4708 & 4168 & 3724 & \\
\end{tabular}
\endgroup
\end{center}
\caption{\small{Confusion matrix.}}
\label{tab:cm3}
\end{table}

\subsection{Higher frequency monitoring}
\label{app:perfo_jumps}

We present here the performances of the monitoring problem described in Section~\ref{sec:jumps}.
After training, the SVR shows a MAPE of around 32 and the SVC has an accuracy of around 95\% on the testing set. 
The confusion matrix is displayed in Table~\ref{tab:cm} for the 12600 testing pairs. 
As can be seen in the Table, the most frequent errors, representing 3.5\% (435/12600) of the testing set, are oscillating shifts predicted as jumps.

\begin{table}[ht]
\begin{center}
\begingroup
\begin{tabular}{lcccc}
\cline{1-4}
\cline{1-4}
Deviation & \multicolumn{3}{c}{Predicted}  & \\
\cline{2-4}
True value & jump & drift & oscillating shift & \\
\cline{1-4}
jump & 4153 & 11 & 36 & 4200 \\
drift &  65 & 4135 & 0 & 4200 \\
oscillating shift & 435 & 10 & 3755 & 4200\\
\cline{1-4}
& 4653 & 4156 & 3791 & \\
\end{tabular}
\endgroup
\end{center}
\caption{\small{Confusion matrix.}}
\label{tab:cm}
\end{table}

\FloatBarrier
\section{Additional figures}
\label{app:figures}

The developed monitoring method has been applied on the number of spots $N_s$, groups $N_g$ and composites $N_c = N_s+10N_g$ at two different scales for the 278 stations in the database. That represents a large amount of information that has not been completely analyzed yet. 
In Section \ref{sec:results}, we present results for typical in-control (IC) and out-of-control (OC) stations, for analyzing the high-frequency as well as low frequency deviations in $N_c$. The station FU is also analyzed at two different scales. 
In this appendix, we present more results about some prominent deviations that occurred in $N_c$.  The persisting drifts are displayed for $\hat \mu_2$ smoothed on a year in Subsection \ref{app:fig_drifts} while the high-frequency deviations are shown for $\hat \mu_2$ smoothed on 27 days in Subsection \ref{app:fig_jumps}. Figures where the monitoring is applied on $N_s$ and $N_g$ in FU are also presented in Subsection \ref{app:fig_Ns_Ng}. \\
The figures are composed of four panels: the upper panel exposes the $\widehat{eh}(i,t)$ of (\ref{E:e2}) for a particular station,
the second panel shows the residuals defined in (\ref{E:residuals}) for the station, the third panel represents the CUSUM statistics applied on the residuals in square-root scale whereas the lower panel displays the characteristics (magnitude and shape) of the shifts predicted by the support vector machine procedures. 

\subsection{Lower frequency monitoring}
\label{app:fig_drifts}

\begin{figure}[!htb]
	\centering
	\begin{subfigure}{0.48\textwidth}
		\centering
		\includegraphics[scale=0.48]{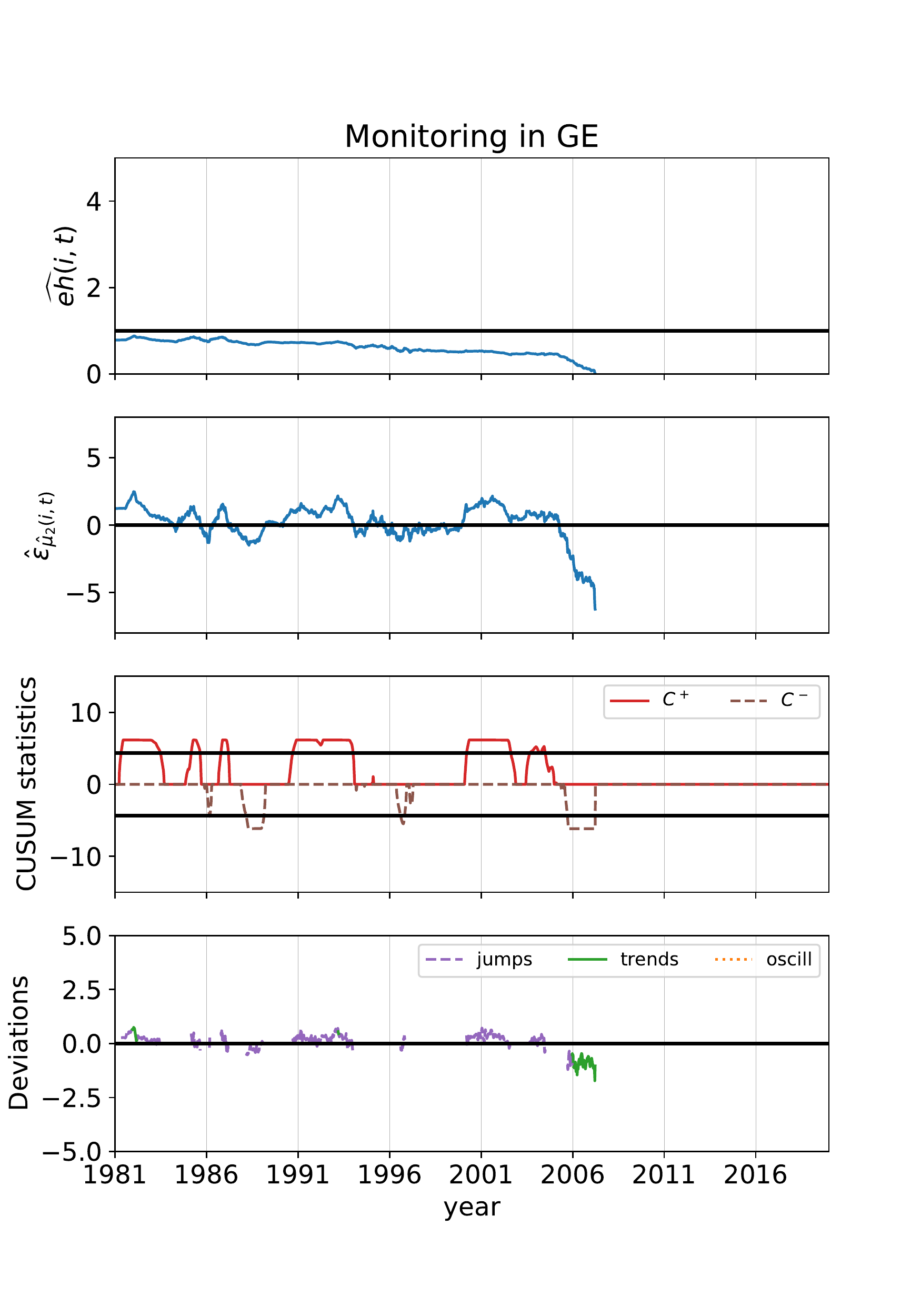}
		\caption{}
		\label{fig:drifts_GE}
	\end{subfigure}
	\begin{subfigure}{0.48\textwidth}
		\centering
		\includegraphics[scale=0.48]{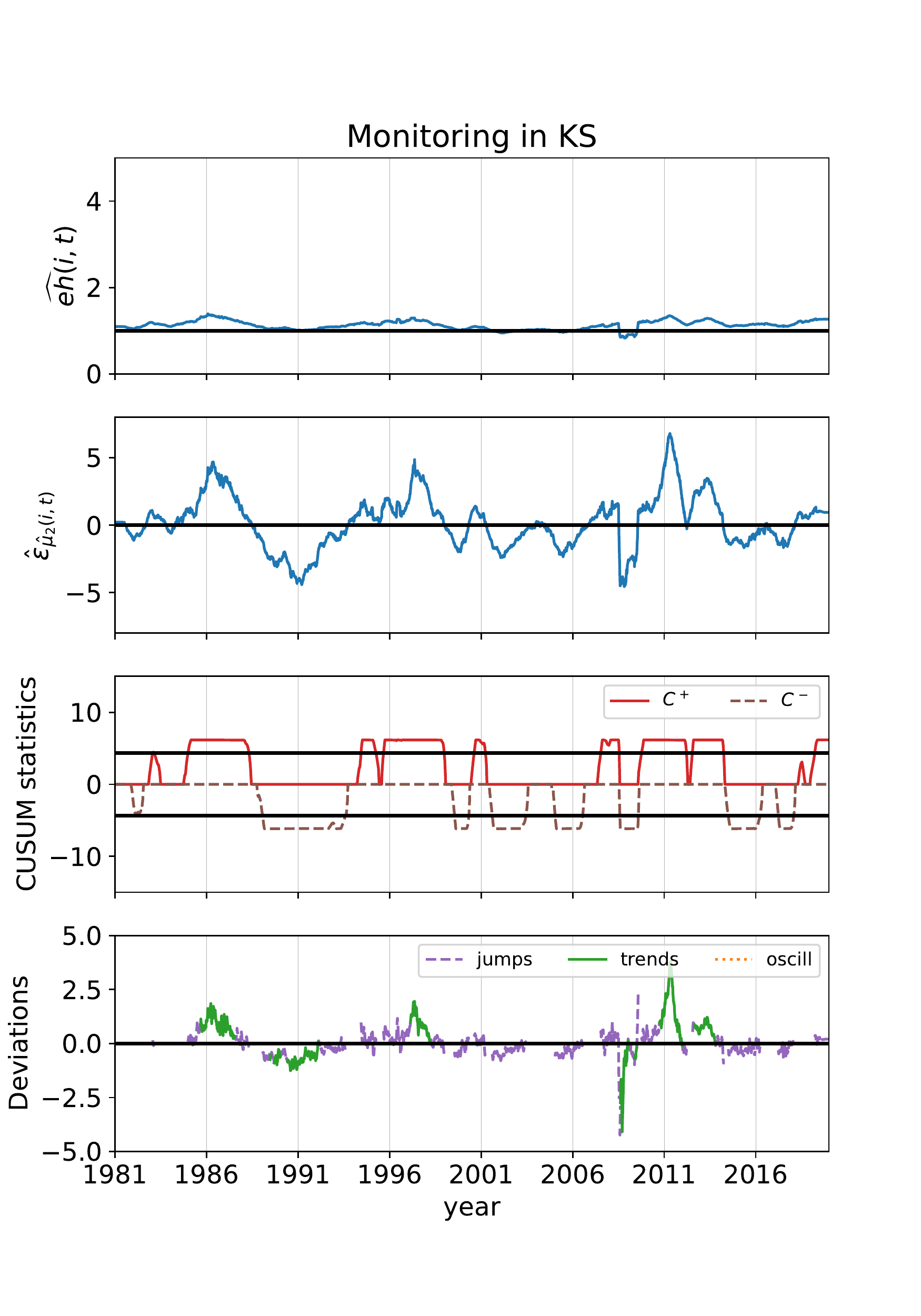}
		\caption{}
		\label{fig:drifts_KS}
	\end{subfigure}
\caption{\footnotesize{ The control scheme applied on the station GE in Belgium over the period studied (1981-2019). b) Similar figure for the station Kislovodsk (KS) in Russia over the same period.}}
\label{fig:drifts_1}
\end{figure}

\begin{figure}[!htb]
	\centering
	\begin{subfigure}{0.48\textwidth}
		\centering
		\includegraphics[scale=0.48]{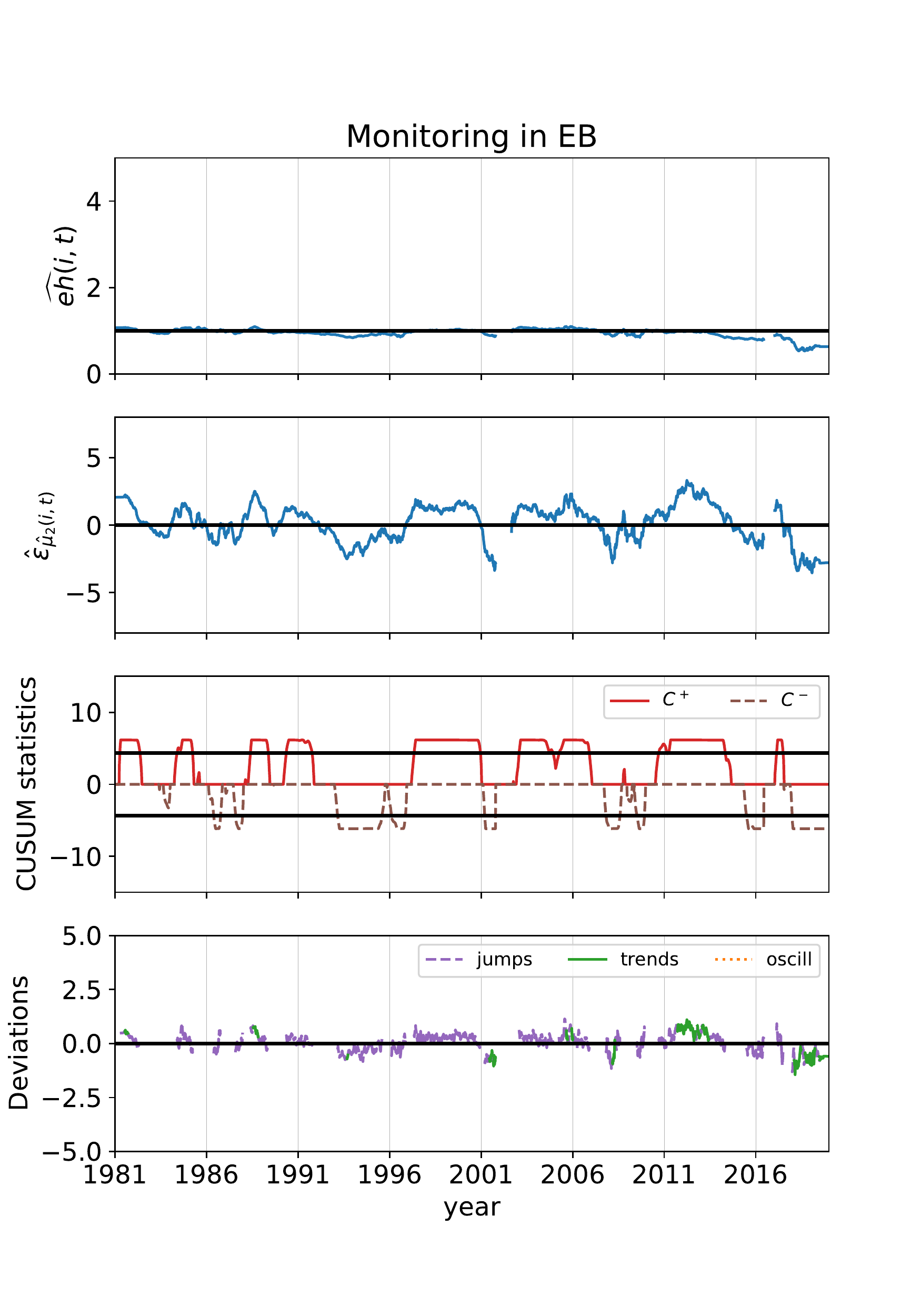}
		\caption{}
		\label{fig:drifts_EB}
	\end{subfigure}
	\begin{subfigure}{0.48\textwidth}
		\centering
		\includegraphics[scale=0.48]{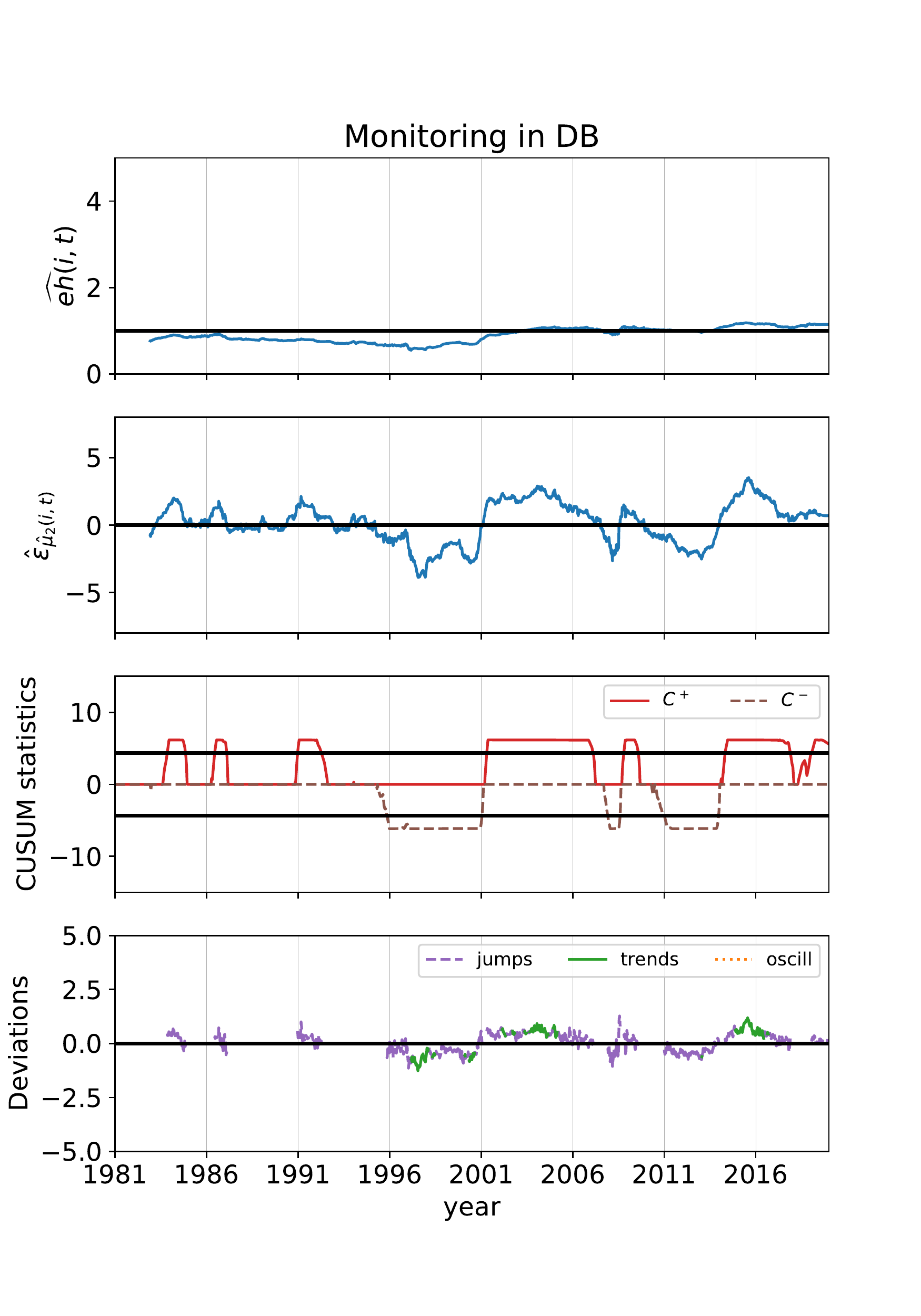}
		\caption{}
		\label{fig:drifts_DB}
	\end{subfigure}
\caption{\footnotesize{The control scheme applied on the data from the Ebro observatory (EB) in Spain over the period studied (1981-2019). b) Similar figure for the station DB in Belgium over the same period.}}
\label{fig:drifts_2}
\end{figure}

In this subsection, we present additional examples of prominent drifts observed in the composite $N_c$ smoothed on a year. 
A downward deviation is visible at the end of the data from the station GE in Figure \ref{fig:drifts_1}. The station GE was composed of a single dedicated observer living in Belgium. The deviation observed in the series is caused by a change of location since the observer moved to another residence at the end of the observations. 
The eyesight of the observer may also have declined with time, which may explain the drop in the series as well. A slight decrease is also visible in the station Kislovodsk (KS) in Russia from mid 2008 till mid 2009. This temporary downward drift may be linked to the installation of a CCD camera in 2008. In 2010, the station also switches to digital image processing, which may cause the slight increase observed afterward. 
In Figure \ref{fig:drifts_2}, a deviation appears in the end of the data from the Ebro observatory (EB) in Spain, which is otherwise very stable. It is probably caused by a change in the camera, which led to problems with the subsequent image treatment. The software was able to process the images again after enhancing the contrast, at the expense of the smallest details in the images. This may probably explain the decrease that we observe at the end of the series. In 1984, the single-observer station DB in Belgium switched from direct observation of the Sun to a projection method that enables the precise drawing of the sunspots. This triggers the small downward deviation observed in the second half of the 1990s since the smallest spots were harder to see with the new procedure. In 2000, the observer bought a new filter, which allowed him to better see the smallest spots. The observer also had more personal time to carefully count the spots and groups. These changes cause a rapid upward shift around 2000-2001 that brings the level of the station back, aligned with those of the network. 

\subsection{Higher frequency monitoring}
\label{app:fig_jumps}

\begin{figure}[!hbt]
	\centering
	\begin{subfigure}{0.48\textwidth}
		\centering
		\includegraphics[scale=0.48]{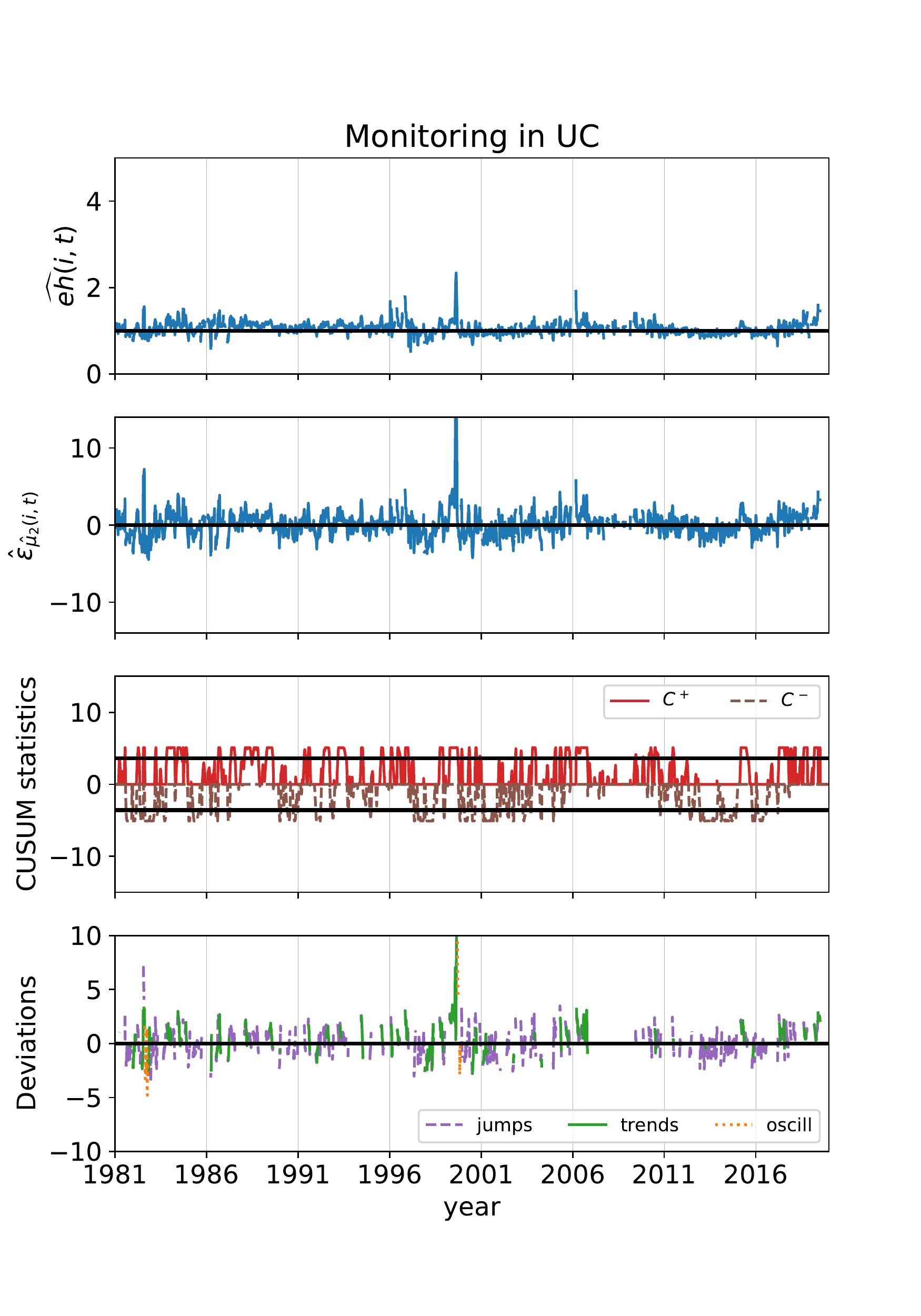}
		\caption{}
		\label{fig:jumps_UC}
	\end{subfigure}
	\begin{subfigure}{0.48\textwidth}
		\centering
		\includegraphics[scale=0.48]{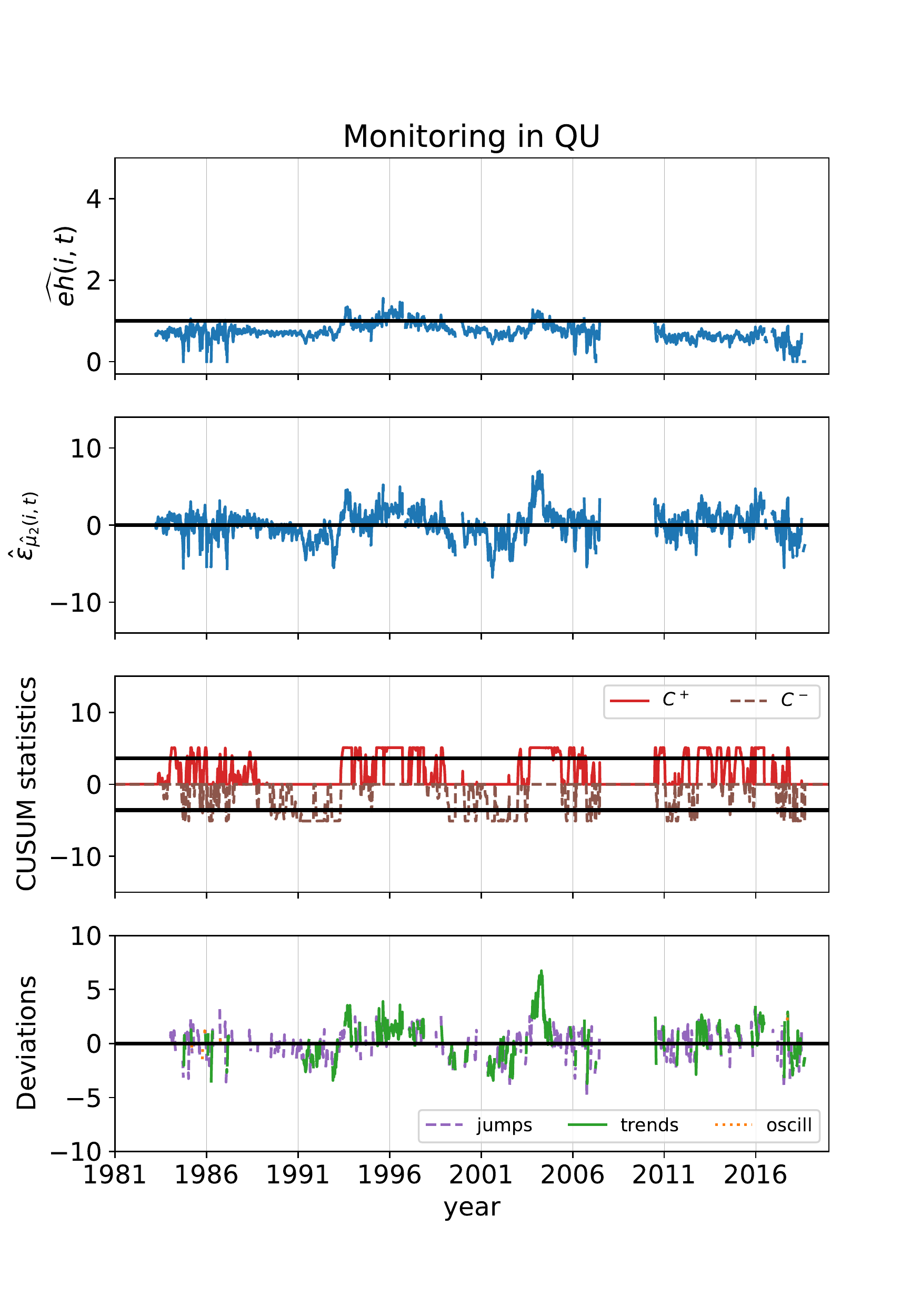}
		\caption{}
		\label{fig:jumps_QU}
	\end{subfigure}
\caption{\footnotesize{ The control scheme applied on the data from the observatory of Uccle (UC) in Belgium over the period studied (1981-2019). b) Similar figure for the station Quezon (QU) in Philippines over the same period.}}
\label{fig:jumps_1}
\end{figure}

\begin{figure}[!htb]
	\begin{subfigure}{0.48\textwidth}
		\centering
		\includegraphics[scale=0.48]{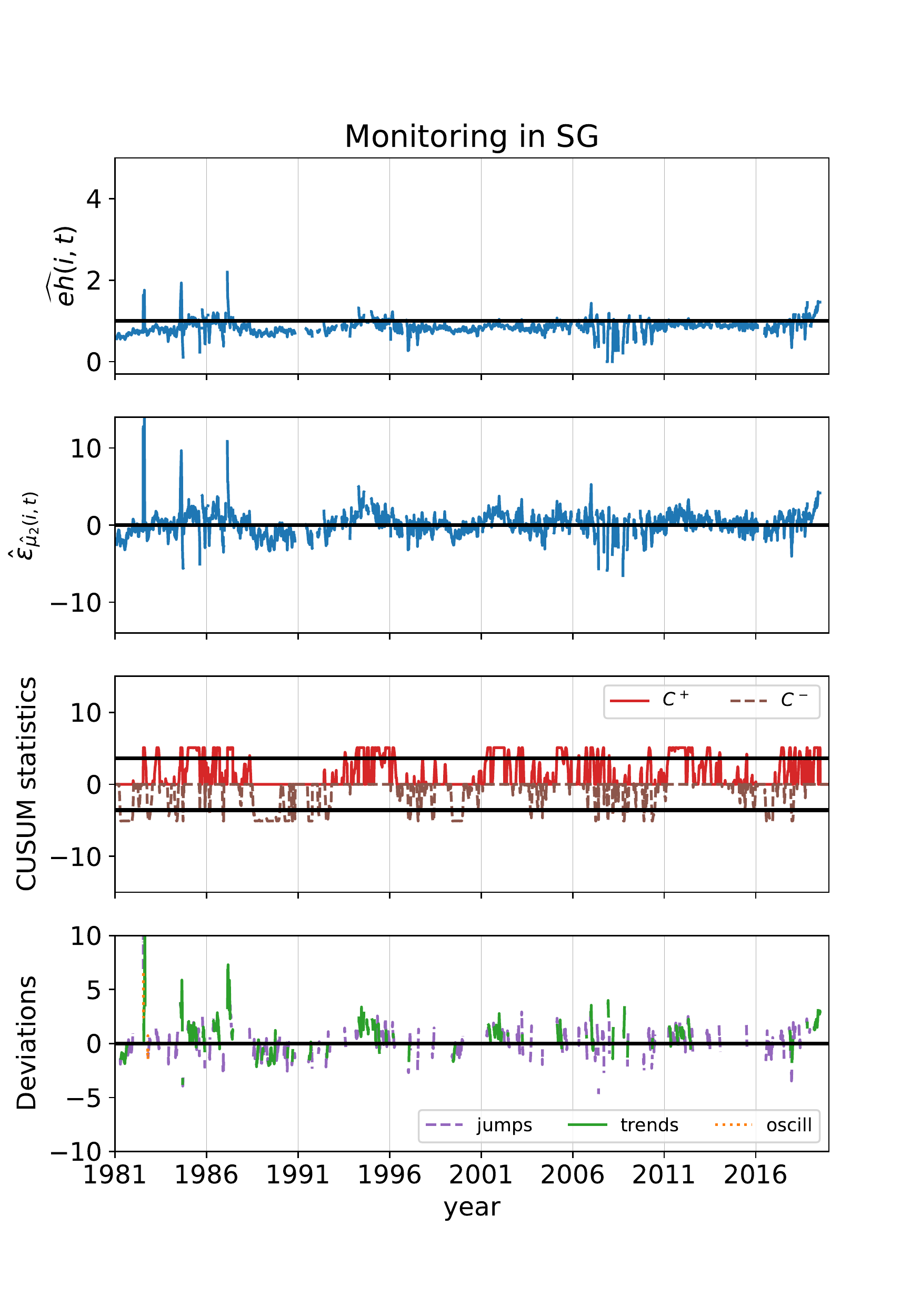}
		\caption{}
		\label{fig:jumps_SG}
	\end{subfigure}
		\centering
	\begin{subfigure}{0.48\textwidth}
		\centering
		\includegraphics[scale=0.48]{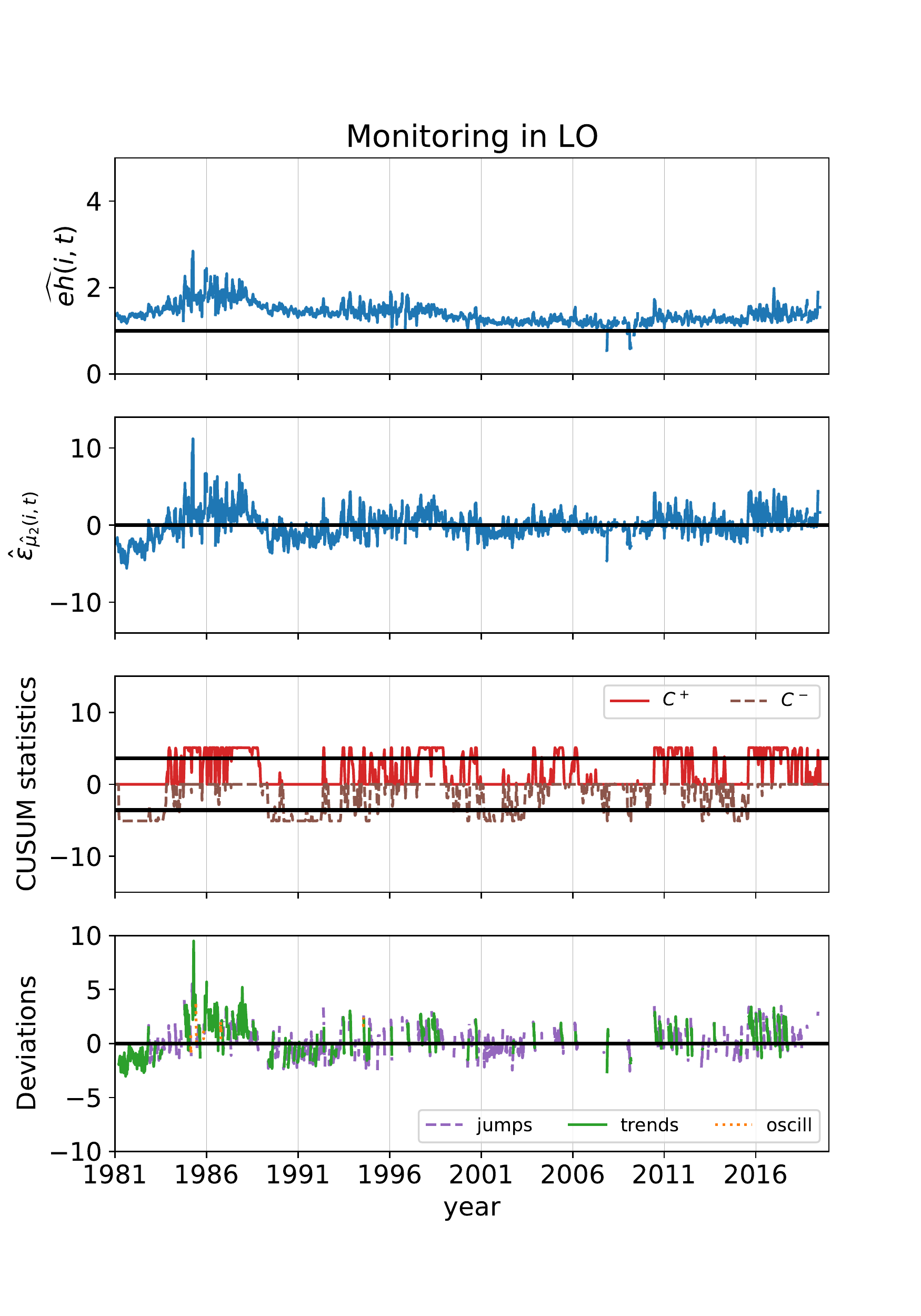}
		\caption{}
		\label{fig:jumps_LO}
	\end{subfigure}
\caption{\footnotesize{ The control scheme applied on the data from the Southern Cross Observatory (SG) in Bolivia over the period studied (1981-2019). b) Similar figure for the observatory of Locarno (LO) in Switzerland over the same period.}}
\label{fig:jumps_2}
\end{figure}

We present here typical examples of major jumps that occurred in the composite $N_c$ smoothed on 27 days.
Figure \ref{fig:jumps_1} shows the method applied on the data from the station Uccle (UC), in Belgium. UC is stable over time but suffers however from a large jump in 1999, already visible in previous analysis \citep{slides_frederic}. This deviating episode is related to the intense participation of a particular observer that did not count with the same precision than the other members of the team. He was recruited at a time where there was a lack of observers but finally stopped observing after a while. The station Quezon (QU) in Philippines experiences more variations over time. They are likely caused by an alternation of observers as in the station San-Miguel (SM) presented in Section \ref{sec:jumps}. 
Figure \ref{fig:jumps_2} displays another type of deviations. The observer from the Southern Cross Observatory (SG) in Bolivia counts more spots than the network at the solar minima (around 1986, 1996 and 2008). This is particularly visible in the three spikes that appear at the beginning of the period studied, soon after he starts observing. Since similar deviations do not happen later in the series, the spikes are most probably related to the learning curve of the observer. We observe later around 1996 and 2008 slight excesses of spots that are also visible in other stations. They may be linked to particular errors that only occur at solar minima when there are few or even no spots on the Sun: some observers tend to over-scrutinize the Sun and may count one or two spots in excess. This effect is not observed in other parts of the solar cycles, when the sunspots are more numerous. Similar deviations are also visible in the observations of the station Locarno (LO) in Switzerland around the solar minima of 1986 (cycle 22). 

\subsection{Results for the number of spots and groups}
\label{app:fig_Ns_Ng}

\begin{figure}[!htb]
	\centering
	\begin{subfigure}{0.48\textwidth}
		\centering
		\includegraphics[scale=0.48]{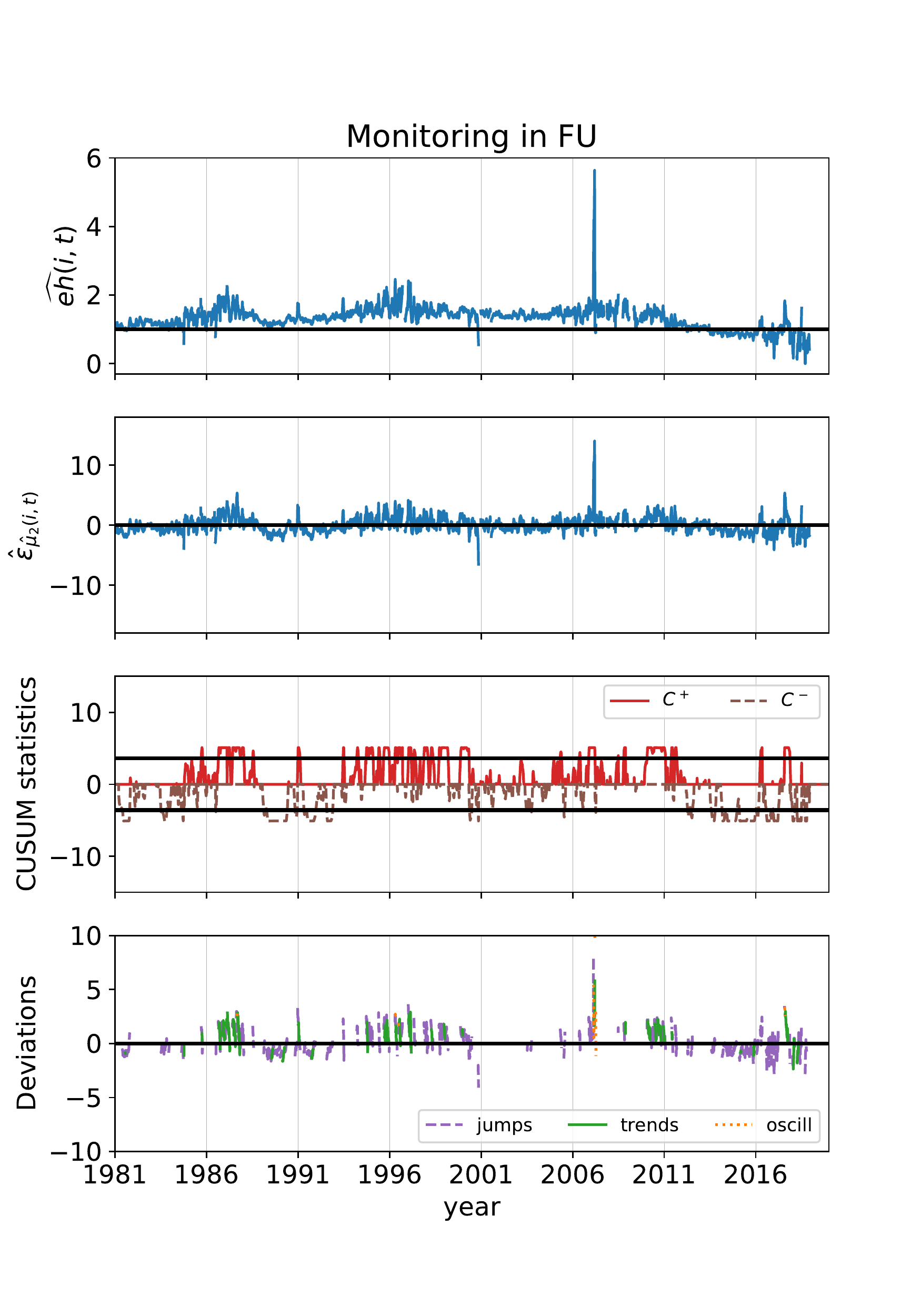}
		\caption{}
		\label{fig:jumps_FU_Ns}
	\end{subfigure}
	\begin{subfigure}{0.48\textwidth}
		\centering
		\includegraphics[scale=0.48]{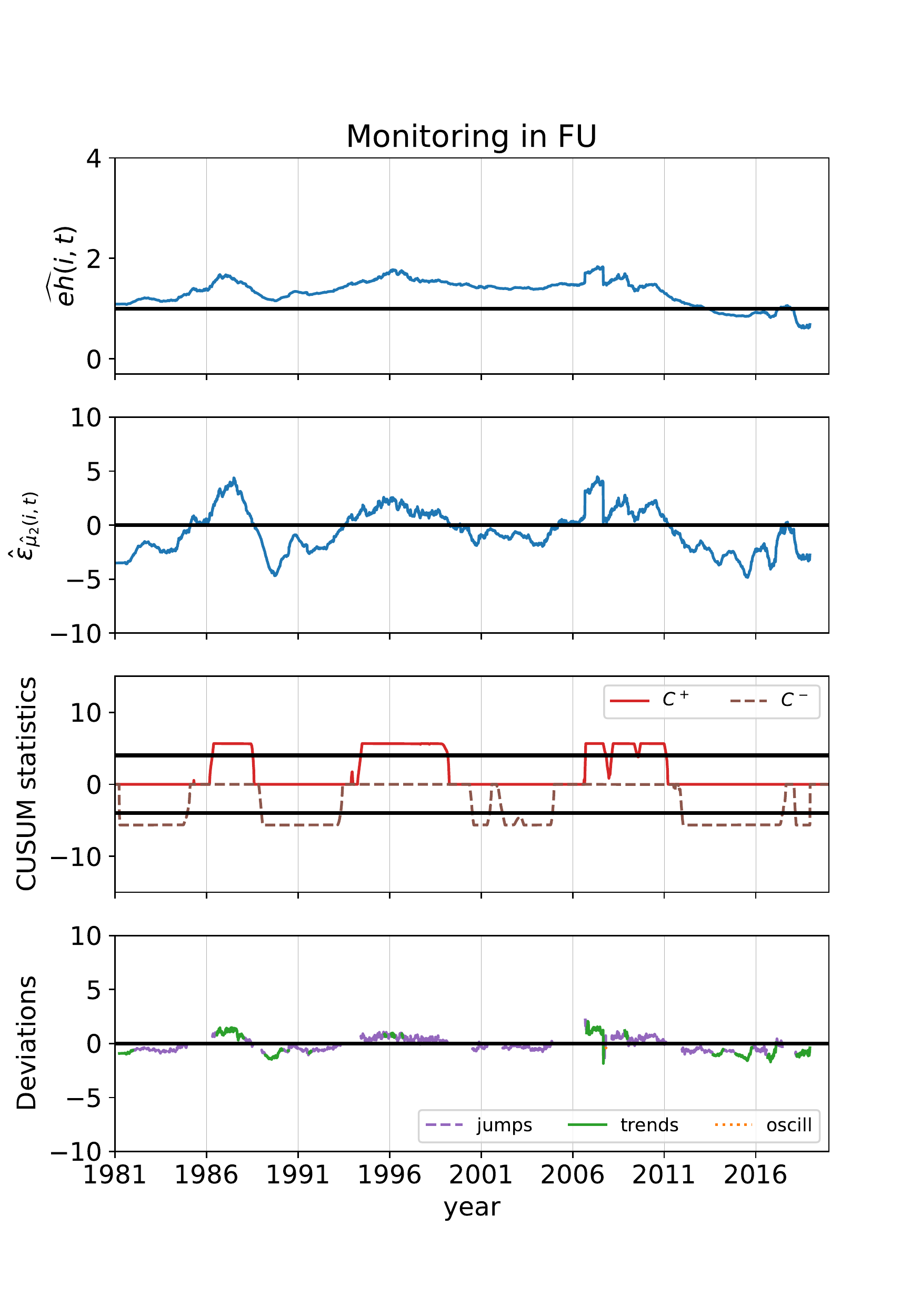}
		\caption{}
		\label{fig:drifts_FU_Ns}
	\end{subfigure}
\caption{\footnotesize{The control scheme applied on the number of spots smoothed on 27 days from station FU in Japan over the period studied (1981-2019). b) Similar figure for $N_s$ smoothed on 365 days in FU over the same period. }}
\label{fig:FU_Ns}
\end{figure}

\begin{figure}[!htb]
	\centering
	\begin{subfigure}{0.48\textwidth}
		\centering
		\includegraphics[scale=0.48]{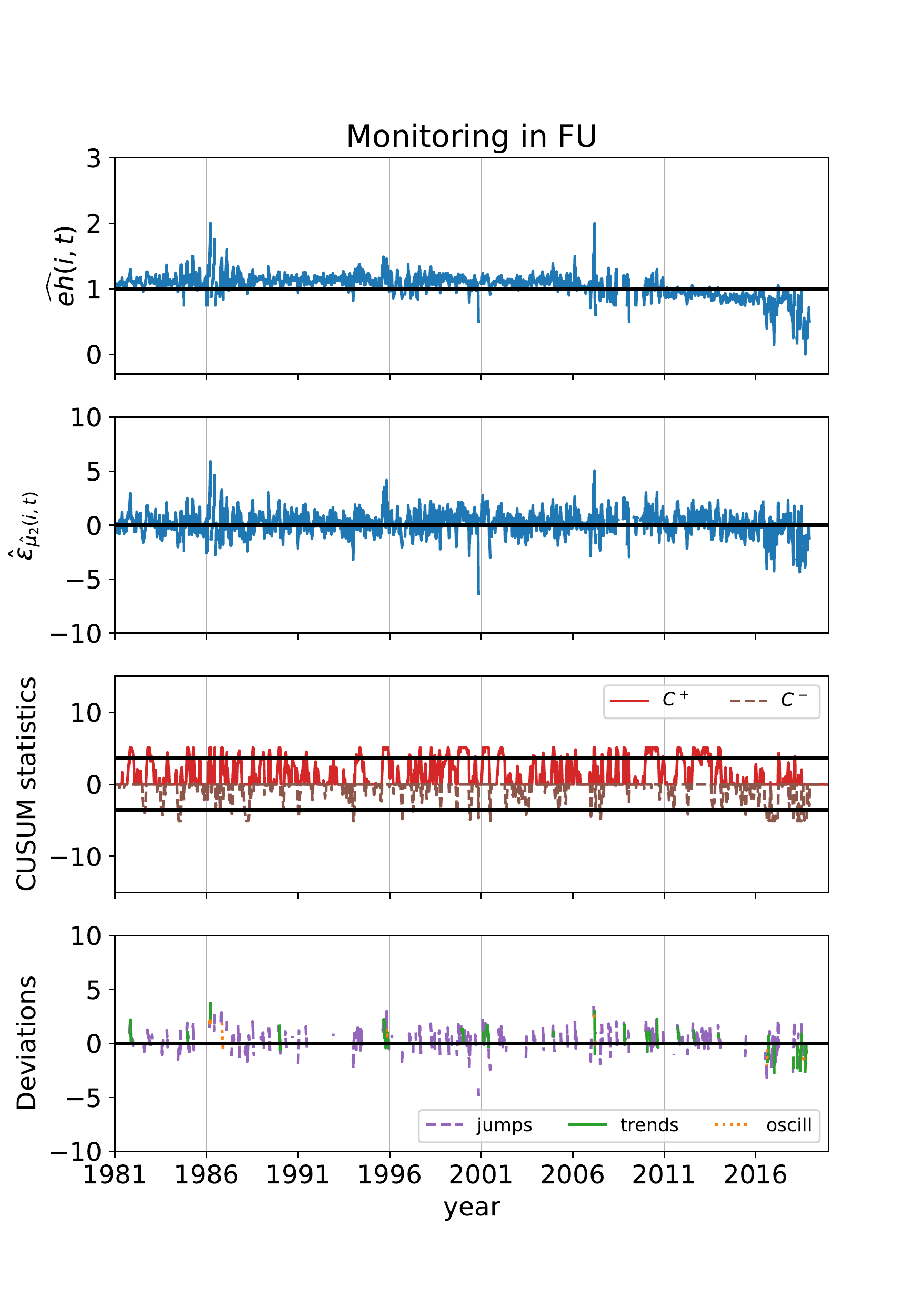}
		\caption{}
		\label{fig:jumps_FU_Ng}
	\end{subfigure}
	\begin{subfigure}{0.48\textwidth}
		\centering
		\includegraphics[scale=0.48]{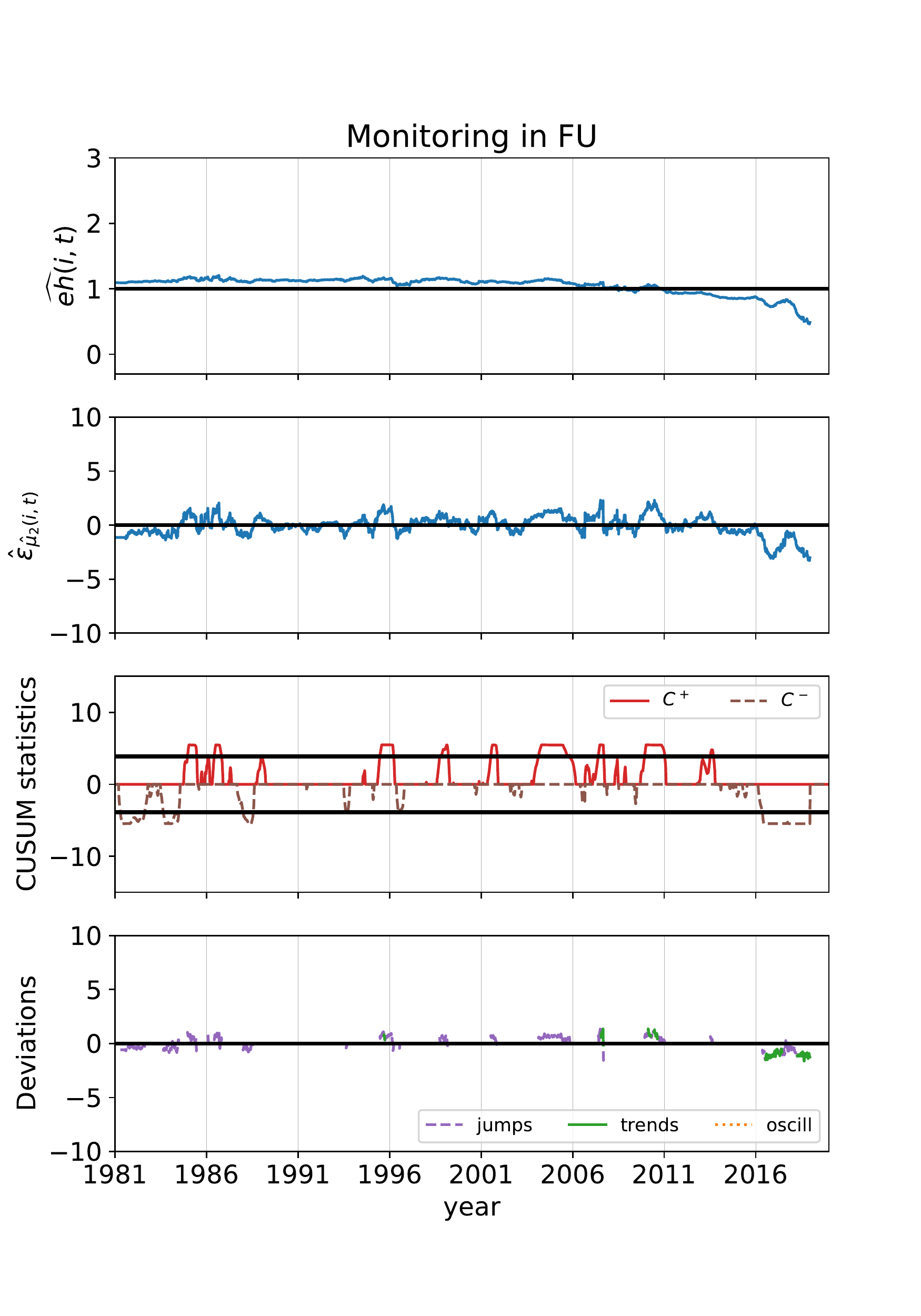}
		\caption{}
		\label{fig:drifts_FU_Ng}
	\end{subfigure}
\caption{\footnotesize{The control scheme applied on the number of groups smoothed on 27 days from station FU in Japan over the period studied (1981-2019). b) Similar figure for $N_g$ smoothed on 365 days in FU over the same period.  }}
\label{fig:FU_Ng}
\end{figure}

The monitoring method has been applied to the data from station FU in Figure \ref{fig:FU} for $N_c$.  Same results are displayed here in Figure \ref{fig:FU_Ns} for the number of spots and in Figure \ref{fig:FU_Ng} for the groups.
As stated in Section \ref{sec:results}, FU is a stable station which is included in the pool. It is presented here since the few deviations of FU are particularly visible and composed of two different types: short-time and persisting shifts. 
As can be seen in the figures, the developed procedure can cope with the different distributions and autocorrelation structures of the data and produces coherent results for $N_s$, $N_g$ as well as $N_c$. \\
Since $N_s$, $N_g$ and $N_c$ are counted on the same image of the Sun captured at a particular moment of the day, similar deviating patterns are visible in the three quantities. The deviations are usually more apparent in $N_s$ than in $N_g$ or $N_c$ since the groups are more robust to counting errors than the individual spots. The jump that occurs in 2007 is for instance much lower in $N_g$ and $N_c$ than in $N_s$. Note that the results presented in the paper focus on $N_c$, which is closer to the International Sunspot Number. \\
By looking separately at $N_s$ and $N_g$, we may also gather more information about the deviations. The drift that appears at the end of the series is gradual in $N_s$ and steep in $N_g$. It may express the fact that the observer progressively sees less spots, which after a certain time leads to a decrease in the number of groups as well. 
Further researches are needed to see if some types of deviations in $N_c$ may be related to rather $N_s$ than $N_g$ or conversely.

\FloatBarrier
\bibliographystyle{apalike}
\bibliography{thesis_biblio}